\documentclass[aps, prd,letterpaper,showpacs,showkeys,twocolumn,floatfix,nofootinbib,superscriptaddress]{revtex4-1} 

\usepackage{amsmath,amssymb,amsfonts,graphicx}
\usepackage{bm}
\usepackage{slashed}
\usepackage[caption=false]{subfig}
\usepackage[svgnames]{xcolor}

\def\a{\alpha}

\def\g{\gamma}

\def\ve{\varepsilon}

\def\lf{\left}
\def\rg{\right}
\def\la{\langle}
\def\ra{\rangle}
\def\no{\nonumber}
\def\hs{\hspace}
\def\ol{\overline}

\newcommand{\vect}[1]{\boldsymbol{#1}}

\def\mathbi#1{\textbf{\em #1}}

\begin{document}

\preprint{ADP-11-34/T756}

\title{Transverse Momentum Dependent Fragmentation and Quark Distribution Functions from the NJL-jet Model}

\author{Hrayr~H.~Matevosyan}
\affiliation{CSSM and ARC Centre of Excellence for Particle Physics at the Tera-scale,\\ 
School of Chemistry and Physics, \\
University of Adelaide, Adelaide SA 5005, Australia
\\ http://www.physics.adelaide.edu.au/cssm
}

\author{Wolfgang Bentz}
\affiliation{Department of Physics, School of Science,\\  Tokai University, Hiratsuka-shi, Kanagawa 259-1292, Japan
\\ http://www.sp.u-tokai.ac.jp/
}

\author{Ian~C.~Clo\"et}
\affiliation{Department of Physics, University of Washington, Seattle WA 98195, USA
\\ http://www.phys.washington.edu/
} 
\affiliation{CSSM and ARC Centre of Excellence for Particle Physics at the Tera-scale,\\ 
School of Chemistry and Physics, \\
University of Adelaide, Adelaide SA 5005, Australia
\\ http://www.physics.adelaide.edu.au/cssm
}

\author{Anthony~W.~Thomas}
\affiliation{CSSM and ARC Centre of Excellence for Particle Physics at the Tera-scale,\\ 
School of Chemistry and Physics, \\
University of Adelaide, Adelaide SA 5005, Australia
\\ http://www.physics.adelaide.edu.au/cssm
}

\begin{abstract}
Using the model of Nambu and Jona-Lasinio to provide a microscopic description 
of both the structure of the nucleon and of the quark to hadron elementary 
fragmentation functions, we investigate the transverse momentum dependence 
of the unpolarized quark distributions in the nucleon and of the quark to 
pion and kaon fragmentation functions. The transverse momentum dependence of the fragmentation functions is
determined within a Monte Carlo framework, with the notable result that the 
average $P_\perp^2$ of the produced kaons is significantly larger than that of
the pions. We also find that $\langle P_\perp^2 \rangle$ has a sizable $z$ dependence,
in contrast with the naive Gaussian ansatz for the fragmentation functions. Diquark
correlations in the nucleon give rise to a nontrivial flavor dependence in the
unpolarized transverse-momentum-dependent quark distribution functions. The 
$\langle k_T^2 \rangle$ of the quarks in the nucleon are also found to have
a sizable $x$ dependence. Finally, these results are used as input to a Monte Carlo
event generator for semi-inclusive deep inelastic scattering (SIDIS), which is used to determine
the average transverse momentum squared of the produced hadrons measured in SIDIS, namely, $\langle P_T^2 \rangle$.
Again, we find that the average $P_T^2$ of the produced kaons in SIDIS is significantly 
larger than that of the pions and in each case $\la P_T^2 \ra$ has a sizable $z$ dependence.
\end{abstract}

\maketitle

\section{Introduction}
\label{SEC_INTRO} 
Semi-inclusive deep inelastic scattering (SIDIS) has a very rich structure which
provides a wealth of observables far in excess of the familiar inclusive 
deep inelastic scattering (DIS). The 2-dimensional picture of a target provided
by SIDIS promises many new insights into nucleon and nuclear 
structure~\cite{Collins:1977iv, Ralston:1979ys,Collins:1984kg,Mulders:1995dh}. 
For example,
it has been realized that SIDIS may shed light on the angular momentum structure of 
the proton in terms of the spin and orbital angular momentum of its quarks and 
gluons~\cite{Bacchetta:2011gx,Avakian:2010br,She:2009jq}.
It will also provide new information on the in-medium modification of bound nucleons
and deepen our understanding of QCD itself~\cite{Collins:1977iv, Ralston:1979ys,Collins:1984kg,Mulders:1995dh}.
The study of the transverse momentum distribution of hadrons 
produced in SIDIS~\cite{Collins:1977iv, Ralston:1979ys,Collins:1984kg,Mulders:1995dh,
Boer:1997nt,Collins:2007ph,Bacchetta:2007wc,Amrath:2005gv} is characterized
by determining the transverse-momentum-dependent (TMD) parton distribution functions (PDFs)
and the TMD fragmentation functions.

Early theoretical models of the fragmentation functions have been constructed in 
Refs.~\cite{Bacchetta:2002tk,Pasquini:2011tk,Jakob:1997wg,Kitagawa:2000ji,Yang:2002gh}
and more recently the development of the NJL-jet model~\cite{Ito:2009zc} has
provided a framework
which automatically satisfies the relevant sum rules. Lattice QCD studies of
TMD PDFs are presented in Ref.~\cite{Musch:2010ka} and 
the QCD evolution of TMD PDFs is discussed in Ref.~\cite{Aybat:2011zv}.
Extensive phenomenological data analysis of transverse momentum in distribution 
and fragmentation processes was presented in Ref.~\cite{Schweitzer:2010tt}. 
Considerable experimental work has already been carried out at 
JLab~\cite{Avakian:2003pk,Osipenko:2008rv,Mkrtchyan:2007sr,Avakian:2010ae,Asaturyan:2011mq}, 
HERMES~\cite{Airapetian:2004tw,Airapetian:2009jy,Airapetian:2009ti} and 
COMPASS~\cite{Alexakhin:2005iw,Ageev:2006da,Rajotte:2010ir}, while for an
overview of the future perspectives for this field we refer to the recent
review by Anselmino {\it et al.}~\cite{Anselmino:2011ay}.

In this work, we present the first microscopic calculation of the spin--independent
TMD quark distribution functions in the nucleon and the TMD quark to pion and kaon
fragmentation functions, where none of the parameters are adjusted to TMD data. 
The underlying theoretical framework is the Nambu--Jona-Lasinio (NJL)
model~\cite{Nambu:1961tp,Nambu:1961fr}. While this certainly represents a simplification of QCD, it has many desirable 
properties. For example, it is covariant and respects the chiral symmetry of QCD, including 
its dynamical breaking. Moreover, it describes
the spin and flavor dependence of the nucleon
PDFs, as well as their modification in-medium~\cite{Cloet:2005pp,Cloet:2005rt,Cloet:2006bq}. 
It also
produces transversity quark distributions~\cite{Cloet:2007em} which are in good agreement 
with the empirical distributions extracted by Anselmino \textit{et al.}~\cite{Anselmino:2007fs}.

For the present purpose the recent developments in the NJL-jet model~\cite{Ito:2009zc, Matevosyan:2010hh, Matevosyan:2011ey}, 
which provides a quark-jet description of the fragmentation process using 
elementary fragmentation functions  calculated within the standard NJL model,
are also critical. 
This framework provides a good description of the parametrizations of experimental 
data and has been extended to include vector meson resonances and nucleons as 
fragmentation channels. The use of Monte Carlo methods to calculate 
these fragmentation functions has also been implemented and that development 
allows us to address a wider array of processes within the model, including physical
cross-section calculations.

In Sec.~\ref{sec:njl_jet_tmd}, we present the general formalism for describing 
transverse momentum distributions in SIDIS, including the quark-jet model originally proposed 
by Field and Feynman. The calculation of the elementary, unintegrated fragmentation functions 
in the NJL model, which are the input to the jet model which describes the TMD fragmentation 
functions in quark hadronization, is explained in Sec.~\ref{sec:SEC_SPLITT}. 
Our model for the TMD quark distribution functions in the nucleon is outlined in 
Sec.~\ref{sec:tmdpdfs}, where we also present results for the TMD PDFs.
Results for the TMD fragmentation functions are discussed in Sec.~\ref{sec:results} and 
the average transverse momentum in the SIDIS process, determined using our Monte Carlo event generator,
is discussed in Sec.~\ref{sec:sidis}.
Finally, Sec.~\ref{sec:conclusions} contains a summary and outlook.

\section{Transverse Momentum in the NJL-jet Model}
\label{sec:njl_jet_tmd}

\begin{figure}[tbp]
\centering\includegraphics[width=1.0\columnwidth]{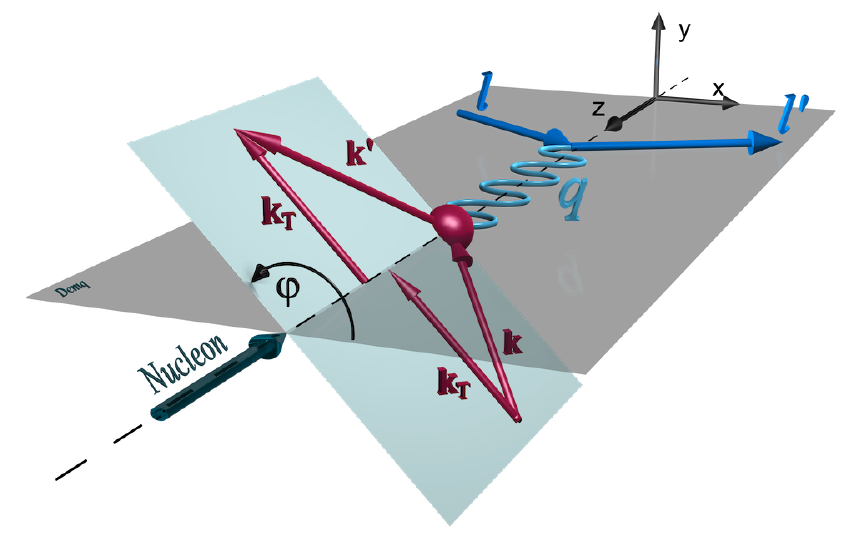}
\caption{Illustration of the three-dimensional kinematics of SIDIS. 
The photon momentum defines the $z$ axis and the struck quark has 
initial transverse momentum $\vect{k_T}$ in the nucleon, with respect to 
the $z$ axis.}
\label{PLOT_DIS_KINEMATICS_3D}
\end{figure}

The kinematics of semi-inclusive hadron production, $lN \to l'hX$, is illustrated schematically 
in Fig.~\ref{PLOT_DIS_KINEMATICS_3D}, where a lepton with momentum $l$ scatters on a target,
by emitting a virtual photon with momentum $q$ that hits a quark with 
initial momentum $k$. As usual, the $z$ axis is chosen to coincide with the
direction of the photon's momentum, where the target has its momentum in the negative
$z$ direction. The transverse momenta in the process -- labeled with a subscript $T$ --
are defined with respect to this $z$ axis, so that the photon and target have no 
transverse momentum component ($\g N$ collinear kinematics).
The angle between the lepton 
scattering plane and the quark scattering plane is denoted as $\varphi$. We allow for 
the struck quark in the target to carry a transverse momentum $\vect{k_{T}}$. 
Some of this transverse momentum is then transferred to the hadrons 
emitted by the quark. 

\begin{figure}[tbp]
\centering\includegraphics[width=1.0\columnwidth]{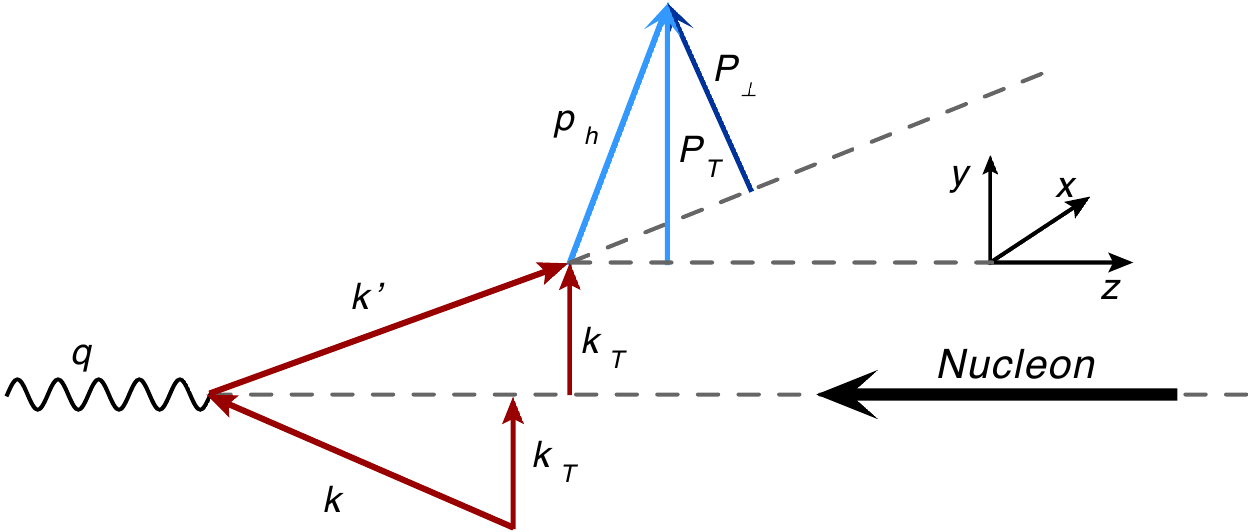}
\caption{Illustration of the kinematics of SIDIS, where the final transverse 
momentum of the produced hadron with respect to the $z$ axis is denoted by $\mathbi{P}_{\mathbi{T}}$, 
which is related to the initial quark transverse momentum in the nucleon $\mathbi{k}_{\mathbi{T}}$ 
and that generated in the fragmentation process $\mathbi{P}_{\mathbf{\perp}}$ by Eq.~\eqref{EQ_PT_PP}.}
\label{PLOT_SIDIS_KIN}
\end{figure}

The kinematics of the quark fragmentation process is depicted 
in Fig.~\ref{PLOT_SIDIS_KIN}. The emitted hadron $h$ carries a 
transverse momentum $\vect{P_T}$ with respect to the $z$ axis which 
can be decomposed into 
two contributions. First, the quark transfers a fraction of its transverse 
momentum $\vect{k_T}$ to the hadron and second the hadron also acquires a momentum 
transverse to the direction of the quark's momentum, $\vect{P_\perp}$. 
Up to corrections of order $\mathcal{O}(k_T^2/Q^2)$, the following relation
holds~\cite{Anselmino:2005nn}:
\begin{equation}
\label{EQ_PT_PP}
\vect{P_T}= \vect{P_\perp} + z\,\vect{k_T} .
\end{equation}
This relation allows one to probe the quark transverse momentum inside a nucleon 
by measuring the $z$ dependence of the emitted hadron's transverse momentum $\langle P_T^2 \rangle$,
provided $\langle P_\perp^2 \rangle$ is independent of $z$. However, in the NJL-jet model framework
we find that $\langle P_\perp^2 \rangle$ is strongly $z$ dependent and this $z$ dependence
is also observed at COMPASS~\cite{Rajotte:2010ir}.
A recent analysis of the HERMES data~\cite{Airapetian:2009jy} was performed in Ref.~\cite{Schweitzer:2010tt},
where a Gaussian ansatz for the TMD quark distribution and fragmentation functions was assumed 
and an average was performed over the quark flavor and type of hadron detected. Using a 
fit region of $0.2<z<0.7$, they extracted the following results for the average transverse 
momentum squared~\cite{Schweitzer:2010tt}:
\begin{align}
\label{EQ_PT_AV_SCHWEITZER}
\langle k_T^2 \rangle = 0.38\pm 0.06\ \mathrm{GeV}^2,\\  
\langle P_\perp^2 \rangle  =  0.16\pm 0.01\ \mathrm{GeV}^2.
\end{align}

\begin{figure}[tbp]
\centering\includegraphics[width=1.0\columnwidth]{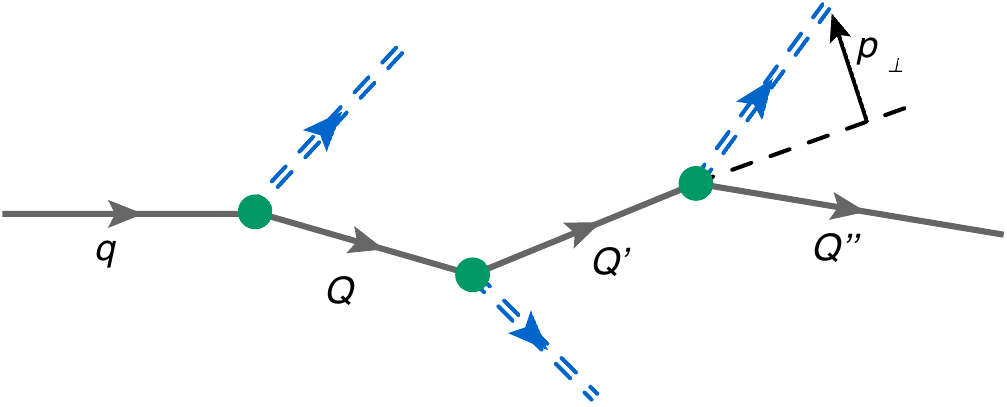}
\caption{NJL-jet model including transverse momentum.}
\label{PLOT_NJL-JET_TMD}
\end{figure}

The latest iteration of the NJL-jet model~\cite{Matevosyan:2011ey} employs 
Monte Carlo simulations to calculate the integrated quark fragmentation functions. 
It assumes that the initial high energy quark emits hadrons in a cascade-like 
process, schematically depicted in Fig.~\ref{PLOT_NJL-JET_TMD}. At every emission 
vertex we choose the type of emitted hadron $h$ and its fraction of the light-cone 
momentum $z$ of the fragmenting quark, by randomly sampling the corresponding 
elementary quark fragmentation (splitting) functions, $\hat{d}_q^h(z)$, that are 
calculated within the NJL model.  
In each elementary fragmentation process we record the flavor of the initial
and final quarks and the type of the emitted hadron, we also note the light-cone 
momentum fraction of the initial quark transferred to the hadron and that left
to the final quark.
The fragmentation chain is stopped after the quark has emitted a predefined number of 
hadrons, $N_{Links}$. We repeat the calculation $N_{Sims}$ times, with the same 
initial quark flavor, $q$, until we have sufficient statistics for the emitted 
hadrons. The fragmentation functions are then extracted by calculating the average 
number of hadrons of type $h$, with light-cone momentum fraction $z$ 
to $z + \Delta z$, which we denote by $\left<N_q^h(z, z+ \Delta z) \right>$.
The fragmentation function in the domain $[z,z+\Delta z]$ is then given by
\begin{equation}
\label{EQ_FRAG_MC}
D_q^h(z) \Delta z = \left< N_q^h(z, z+ \Delta z) \right> \equiv  \frac{ \sum_{N_{Sims}} N_q^h(z, z+ \Delta z) } { N_{Sims} }.
\end{equation}

In this work, we extend the NJL-jet model to include the transverse momentum dependence
of the emitted hadrons in the fragmentation process. This is achieved
by using TMD elementary quark fragmentation functions at the hadron 
emission vertices and by keeping track of the transverse momenta of all the particles in 
the process. Our goal is to calculate the TMD fragmentation function, 
$D_q^h(z,P_\perp^2)$, using its probabilistic interpretation. That is, the probability of a 
quark $q$ to emit a hadron $h$ with a fraction $z$ of its light-cone momentum 
and a transverse momentum $\vect{P_\perp}$ is given by $D_q^h(z,P_\perp^2)\,dz\, d^2\vect{P}_\perp$.

\begin{figure}[tbp]
\centering\includegraphics[width=1.0\columnwidth]{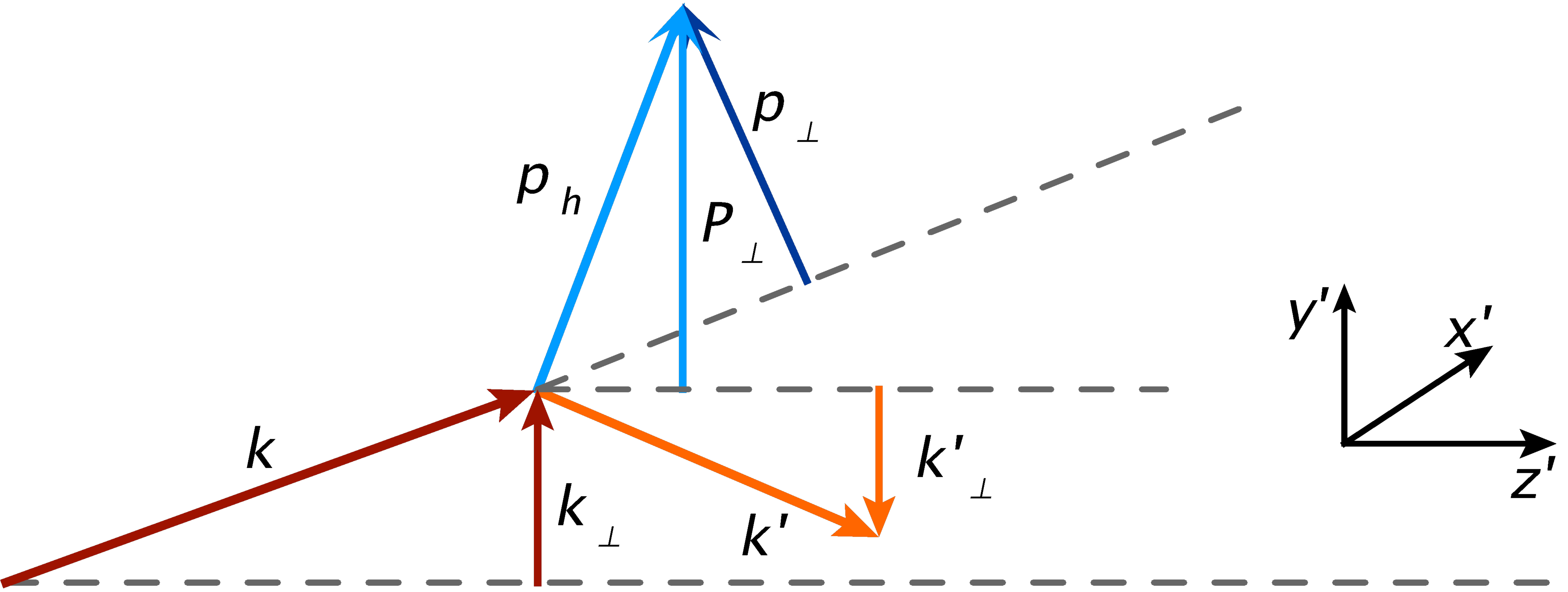}
\caption{Quark elementary fragmentation kinematics, for an arbitrary hadron emission
in the cascade chain. The $z'$ axis is defined by the direction of the 3-momentum
of the original parent quark.}
\label{PLOT_SPLITTING_KIN}
\end{figure}

We calculate elementary (one-step) TMD splitting functions, $\hat{d}_q^h(z,p_\perp^2)$, using the NJL model, 
where $\vect{p_\perp}$ denotes the transverse component of the hadron's momentum with respect 
to the parent quark, as illustrated in Figs.~\ref{PLOT_NJL-JET_TMD} and \ref{PLOT_SPLITTING_KIN}. 
In each step of the Monte Carlo simulation of the quark cascade emission, we randomly sample the type, 
the light-cone momentum fraction, $z$, and the transverse momentum, $\vect{p_\perp}$, of 
the emitted hadron using as the probability distribution the elementary TMD splitting functions of the quark, 
where the elementary probability is $\hat{d}(z,p_\perp^2)\,dz\, d^2\vect{p}_\perp$. Schematically, the quark emission process is 
depicted in Fig.~\ref{PLOT_SPLITTING_KIN}, 
where the $z'$ axis denotes the direction of the original parent quark's 3-momentum. The vectors $\vect{k}$ 
and $\vect{k'}$ denote the 3-momentum of an arbitrary quark in the cascade chain before and after hadron 
emission with transverse components $\vect{k_\perp}$ and $\vect{k'_\perp}$, respectively. 
The emitted hadron's momentum is labeled by $\vect{p_h}$, where its transverse 
component with respect to $\vect{k}$ and the $z'$ axis is denoted by $\vect{p_\perp}$ and $\vect{P_\perp}$,
respectively. 
$\vect{P_\perp}$ is obtained using the relation $\vect{P_\perp}= \vect{p_\perp} + z\,\vect{k_\perp}$, 
analogous to that in Eq.~(\ref{EQ_PT_PP}). 
The recoil transverse momentum of the final quark, $\vect{k'_\perp}$, is calculated from 
momentum conservation in the transverse plane, namely,
\begin{equation}
\label{EQ_PT_QUARK}
\vect{k_\perp}= \vect{P_\perp} + \vect{k'_\perp}.
\end{equation}
The TMD fragmentation function is then calculated after the trivial integration 
over the polar angle of $\vect{P}_\perp$ in the transverse plane, that is
\begin{align}
&D_q^h(z,P_\perp^2)\, \Delta z\, \pi \Delta P_\perp^2 = \left< N_q^h(z, z+ \Delta z,P_\perp^2,P_\perp^2+ \Delta P_\perp^2 \right> \nonumber \\
&\hspace{12mm}
\equiv  \frac{ \sum_{N_{Sims}} N_q^h(z, z+ \Delta z,P_\perp^2,P_\perp^2+ \Delta P_\perp^2)} { N_{Sims} }.
\label{EQ_FRAG_MC_TMD}
\end{align}

The model can easily accommodate the initial transverse momentum of the quark, for 
example, with respect to the direction of the virtual photon in SIDIS 
(see Fig.~\ref{PLOT_DIS_KINEMATICS_3D}).  Our goal is to describe the average 
transverse momentum of the hadrons produced in different reactions. 
The differential cross section for SIDIS up to terms of order 
$\mathcal{O}(k_T^2/Q^2)$ can be written as \cite{Anselmino:2005nn}
\begin{align}
\label{EQ_SIDIS_X_SEC}
\frac{d^3\sigma^{l N \to l' hX}}{dx\ dz\ d P_T^2} &\sim \sum_q e^2_q \int d^2 \vect{k_T}\ q(x, k_T^2) \ D_q^h(z, P_\perp^2) \no \\
&\equiv \sum_q \, e_q^2\,\widetilde{D}_q^h(x,z,P_T^2),
\end{align}
where $\vect{P}_\perp$ and $\vect{P_T}$ are related by Eq.~\eqref{EQ_PT_PP} and $q(x, k_T^2)$ are
the TMD quark distribution functions of the target. Thus for SIDIS, 
we can use the TMD quark distribution functions to randomly sample the initial 
transverse momentum of the quark to calculate the relevant  number density of the produced hadrons. 
In this work, we use TMD valence quark distributions in the nucleon -- calculated within 
the NJL model -- to determine the average transverse 
momentum of the produced hadrons with respect to the direction of the virtual photon, 
that is $\langle P_T^2\rangle$, by calculating the corresponding probability densities
$\widetilde{D}_q^h(x,z,P_T^2)$, using an expression analogous to Eq.~\eqref{EQ_FRAG_MC_TMD}. 
In this way, we obtain a self-consistent description of the entire process in the 
regime where the virtual photon samples the valence quark component of the target,
that is, when the struck quark has $x\gtrsim 0.3$. The type of target and the allowed range 
of $x$ in the Monte Carlo simulation can be matched to those measured in any particular 
experiment. 

In this article, we only consider the production of pseudoscalar mesons, that is, the pions and
kaons, as a first step in determining the TMD fragmentation functions. Eventually, we will 
also include the vector mesons and nucleon-antinucleon channels, as done for the
integrated fragmentation functions in Ref.~\cite{Matevosyan:2011ey}.

\section{Elementary TMD Fragmentation Functions}
\label{sec:SEC_SPLITT}

In this section, we evaluate 
the ``elementary" fragmentation functions of quarks to hadrons as a ``one-step" 
process in the NJL model, using light-cone coordinates.\footnote{We use 
the following LC convention for Lorentz 4-vectors 
$(a^+,a^-,\mathbi{a}_\perp)$, $a^\pm=\frac{1}{\sqrt{2}}(a^0\pm a^3)$ and $\mathbi{a}_\perp = (a^1,a^2)$. }
The NJL model which we use includes only four point quark interactions in the 
Lagrangian, with up, down and strange quarks (see, for example, 
Refs.~\cite{Kato:1993zw,Klimt:1989pm,Klevansky:1992qe} for detailed reviews of 
the NJL model). In the present work, we use the notation introduced in our previous 
studies~\cite{Matevosyan:2010hh,Matevosyan:2011ey}. 

\begin{figure}[bp]
\centering\includegraphics[width=1.0\columnwidth]{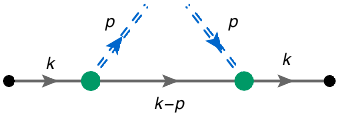}
\caption{Feynman diagram describing the elementary quark to hadron fragmentation functions.}
\label{PLOT_FRAG_QUARK}
\end{figure}

The elementary fragmentation function for quark, $q$, to emit a meson, $m$, 
carrying light-cone momentum fraction, $z$, and carrying transverse momentum, 
$p_\perp$, is depicted in Fig.~\ref{PLOT_FRAG_QUARK}. In the frame where 
the fragmenting quark has zero transverse momentum, but a nonzero transverse momentum 
component $-\vect{p_\perp}/z$ with respect to the direction of the produced hadron~\cite{Collins:1977iv,Ito:2009zc}, 
the unregularized elementary TMD fragmentation functions to pseudoscalar mesons 
are given by
\begin{widetext}
\begin{align}
\label{EQ_QUARK_FRAG_TMD}
\nonumber
d_{q}^{m}(z,p_\perp^2) &= -\frac{C_q^m}{2}\,  g_{mqQ}^{2}\, \frac{z}{2} \int \frac{dk_+ dk_-}{(2\pi)^{4}} 
\ \mathrm{Tr}\lf[S_{1}(k)\gamma^{+}S_{1}(k)\gamma_{5} (\slashed{k}-\slashed{p}+M_{2}) \gamma_{5}\rg]
\delta(k_{-} - p_{-}/z)\ 2\, \pi\, \delta\!\lf( (p-k)^{2} -M_{2}^{2} \rg) \nonumber \\
&= \frac{C_q^m }{16\pi^{3}}\, g_{mqQ}^{2}\, z\  \frac{p_{\perp}^{2} + \lf[(z-1)M_{1}+M_{2}\rg]^{2}} 
{\lf[p_{\perp}^{2}+z(z-1)M_{1}^{2}+zM_{2}^{2}+(1-z)m_{m}^{2}\rg]^{2}}.
\end{align}
\end{widetext}
The trace is over Dirac indices only and the subscripts on the quark propagator, $S_1(k)$, 
and constituent masses, $M_1$ and $M_2$, denote quark flavors. Quark flavor is also indicated by the subscripts $q$ and $Q$, where a meson of type 
$m$ has the quark flavor structure $m=q\overline{Q}$ and $m_m$ denotes the meson mass.
The corresponding isospin factor and quark-meson 
coupling constant are labeled by $C_q^m$ and $g_{mqQ}$, respectively, and are determined 
within the NJL model~\cite{Matevosyan:2010hh,Matevosyan:2011ey}. The integrated elementary splitting 
function is obtained from the elementary TMD splitting function via integration over $\vect{p}_\perp$, that is,
\begin{align}
\label{EQ_QUARK_FRAG}
&d_{q}^{m}(z) = \int d^2 \vect{p}_\perp\  d_{q}^{m}(z,p_\perp^2)
= \frac{C_q^m }{2} g_{mqQ}^{2}\ z \nonumber \\
&\times
\int \frac{d^2 \vect{p}_\perp}{(2\pi)^{3}} 
\frac{p_{\perp}^{2} + \lf[(z-1)M_{1}+M_{2}\rg]^{2}} 
{\lf[p_{\perp}^{2}+z(z-1)M_{1}^{2}+zM_{2}^{2}+(1-z)m_{m}^{2}\rg]^{2}}.
\end{align}
The probability densities $\hat{d}_{q}^{m}(z)$ are then obtained by multiplying
a normalization factor so that $\sum_m \int_0^1dz\, \hat{d}_{q}^{m}(z) = 1$. The isospin
and momentum sum rules are then satisfied automatically~\cite{Matevosyan:2010hh,Matevosyan:2011ey}.

Previously, we employed the Lepage-Brodsky (LB) regularization scheme to calculate loop integrals 
such as that in Eq.~(\ref{EQ_QUARK_FRAG}). This method puts a sharp cutoff on the 
invariant mass squared, $M_{12}^2$, of the particles in the final state 
(see Refs.~\cite{Bentz:1999gx,Ito:2009zc,Matevosyan:2010hh,Matevosyan:2011ey} for a 
detailed description as applied to the NJL-jet model). 
The maximum invariant mass of the two particles in the loop, $\Lambda_{12}$, is 
determined by
\begin{align}
\label{EQ_LB_REG}
M_{12}^2\leq \Lambda_{12}^2 \equiv \left(\sqrt{\Lambda_{3}^{2} + \mu_1^{2}} +\sqrt{\Lambda_{3}^{2} + \mu_2^{2}}\right)^2,
\end{align}
where $\mu_1$ and $\mu_2$ denote the masses of the particles in the loop and
$\Lambda_3$ denotes the 3-momentum cutoff, which is fixed in the usual way by
reproducing the experimental pion decay constant. For a light constituent quark
mass of $M=0.4~\mathrm{GeV}$, the corresponding 3-momentum cutoff is $\Lambda_3= 0.59 ~{\rm GeV}$. 
The strange constituent quark mass is determined by reproducing the experimental kaon mass,
giving the value $M_s=0.61~\mathrm{GeV}$ and the corresponding quark-meson coupling constants 
are $g_{\pi qQ}=4.23$ and $g_{K qQ}=4.51$.

In loop integrals containing two particles, we assign a light-cone momentum fraction
$x$ (of the initial particle's light-cone momentum) to the particle with mass $\mu_1$ 
and consequently a light-cone momentum fraction $1-x$ for the particle with mass $mu_2$.
Then, in the frame where the initial particle's transverse momentum is zero, 
the invariant mass of the two particles in the loop can be expressed as
\begin{eqnarray}
\label{EQ_LB_INV_M}
M_{12}^{2} = \frac{\mu_{1}^{2}+p_{\perp}^{2}}{x} + \frac{\mu_{2}^{2}+p_{\perp}^{2}}{1-x}.
\end{eqnarray}
The relation in Eq.~\eqref{EQ_LB_REG}, when applied to the integral in 
Eq.~(\ref{EQ_QUARK_FRAG}), yields a sharp cutoff in the integral over the 
transverse momentum, namely
\begin{multline}
\label{EQ_QUARK_FRAG_LB_PPERP}
p_\perp^2 \leqslant \mathcal{P}_\perp^2 \equiv 
z(1-z)\left[\sqrt{\Lambda_{3}^2 + \mu_1^2} + \sqrt{\Lambda_{3}^2 + \mu_2^2}\right]^2 \\
- (1-z)\,\mu_1^2 -z\,\mu_2^2.
\end{multline}

\begin{figure}[tbp]
\centering\includegraphics[width=1.0\columnwidth]{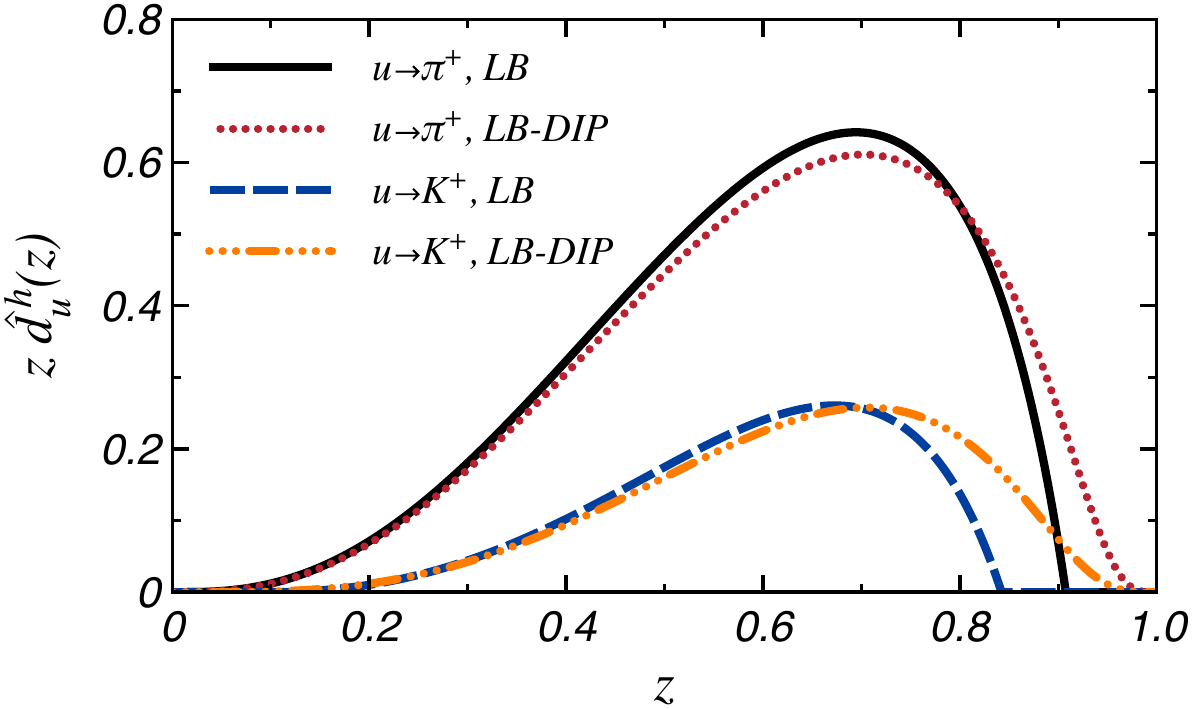}
\caption{The normalized integrated splitting functions for a $u$ quark 
to $\pi^+$ and $K^+$, calculated using LB and LB-DIP regularizations 
with the same light constituent quark mass of $M=0.4\,\mathrm{GeV}$.}
\label{PLOT_DRV_LB_DIP}
\end{figure}

A consequence of LB regularization is that it restricts the corresponding regularized functions 
 to a limited range of $z$, namely
 $0 < z_{min} \leqslant z \leqslant z_{max} < 1$, 
where $z_{min}$ and $z_{max}$ are determined by imposing the 
condition $\mathcal{P}_{\perp}^2\geqslant 0$ in Eq.~(\ref{EQ_QUARK_FRAG_LB_PPERP}). 
These range limitations depend on the masses of hadrons and quarks involved. 
For example, the $z$ limits are very close to the endpoints ($z=0$ and $z=1$) for 
quark splitting functions to pions, but are further from these endpoints for heavier hadrons like kaons. 
The plots depicted in Fig.~\ref{PLOT_DRV_LB_DIP} show the limited range for the 
normalized splitting functions of a $u$ quark to $\pi^+$ and $K^+$, 
calculated using LB regularization.

In this work, we employ a slightly modified version of the LB regularization, 
which replaces the sharp cutoff of the invariant mass squared in the 
integrals, namely, $\Theta(\Lambda_{12}^2-M_{12}^2)$, by a dipole regulator: 
\begin{eqnarray}
\label{EQ_LB_DIP}
G_{12}(p_\perp^2) \equiv \frac{1}{\left[ 1 + (M_{12}^2/\Lambda_{12}^2)^2 \right]^2}.
\end{eqnarray}
A physical motivation for this regularization scheme is that it gives a pion quark 
distribution that at large $x$ behaves approximately as $(1-x)^{2.6}$ for $Q^2 = 16\,$GeV$^2$,
which is in good agreement with the recent reanalysis of Aicher \textit{et al.} which
finds $(1-x)^{2.34}$ at the same scale~\cite{Aicher:2010cb}.
Using this dipole cutoff version of the LB regularization scheme (LB-DIP), we fix the model parameters
by reproducing the experimentally measured hadronic properties, such as $f_\pi$
and the kaon mass to determine the cutoff as $\Lambda_{3}=0.773~\mathrm{GeV}$ and the 
strange constituent quark mass becomes $M_s=0.59~\mathrm{GeV}$. The corresponding 
quark-meson coupling constants are $g_{\pi qQ}=4.24$ and $g_{K qQ}=4.52$. The
quark distribution functions calculated with LB-DIP regularization satisfy both the number and
momentum sum rules and allow us to set the model scale at $Q_0^2=0.2~\mathrm{GeV^2}$ in the 
usual way by comparing the evolved pion distribution function with that obtained 
from experiment. This procedure is discussed in detail in Ref.~\cite{Matevosyan:2010hh}.
 
\begin{figure}[tbp]
\subfloat{\centering \includegraphics[width=1.0\columnwidth]{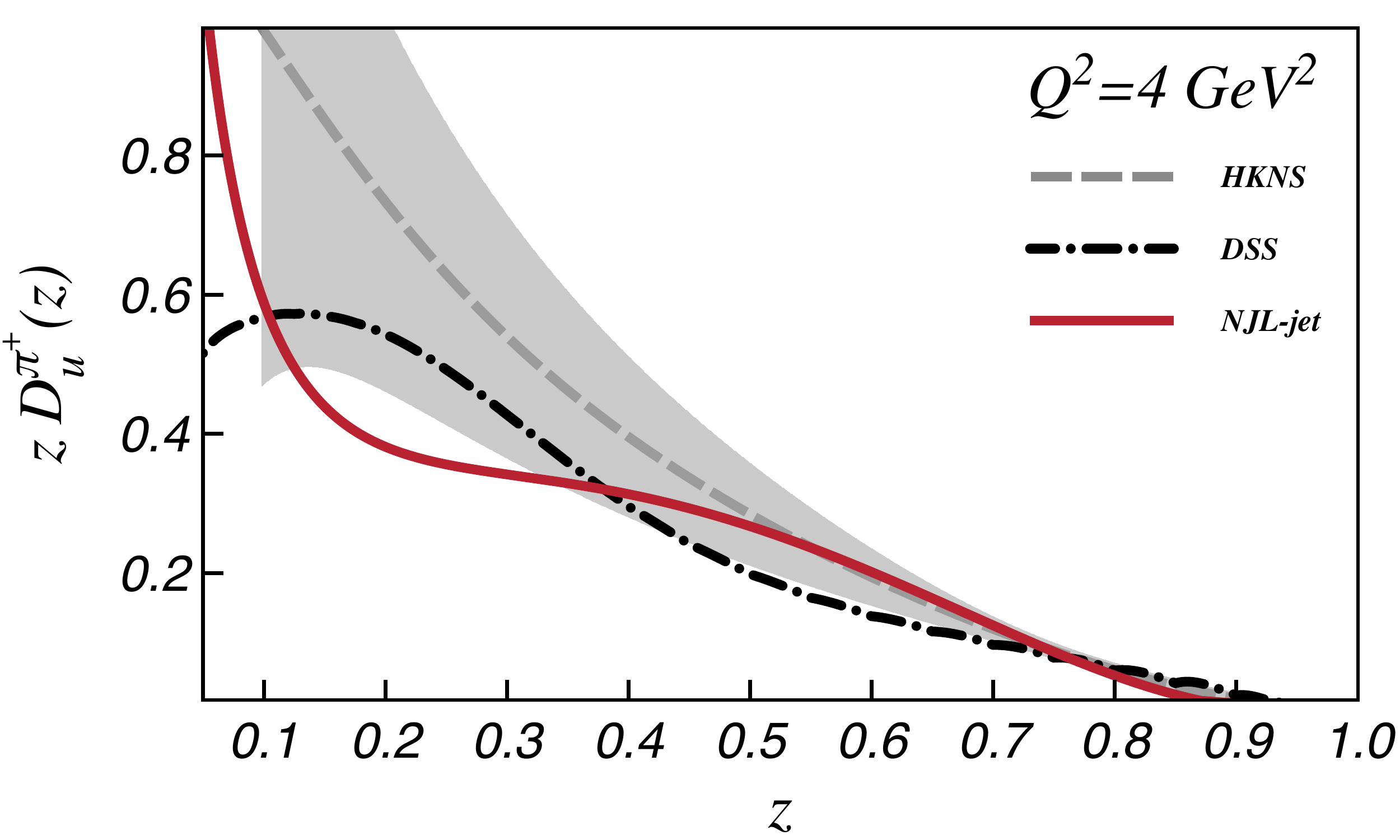}} \\
\subfloat{\centering \includegraphics[width=1.0\columnwidth]{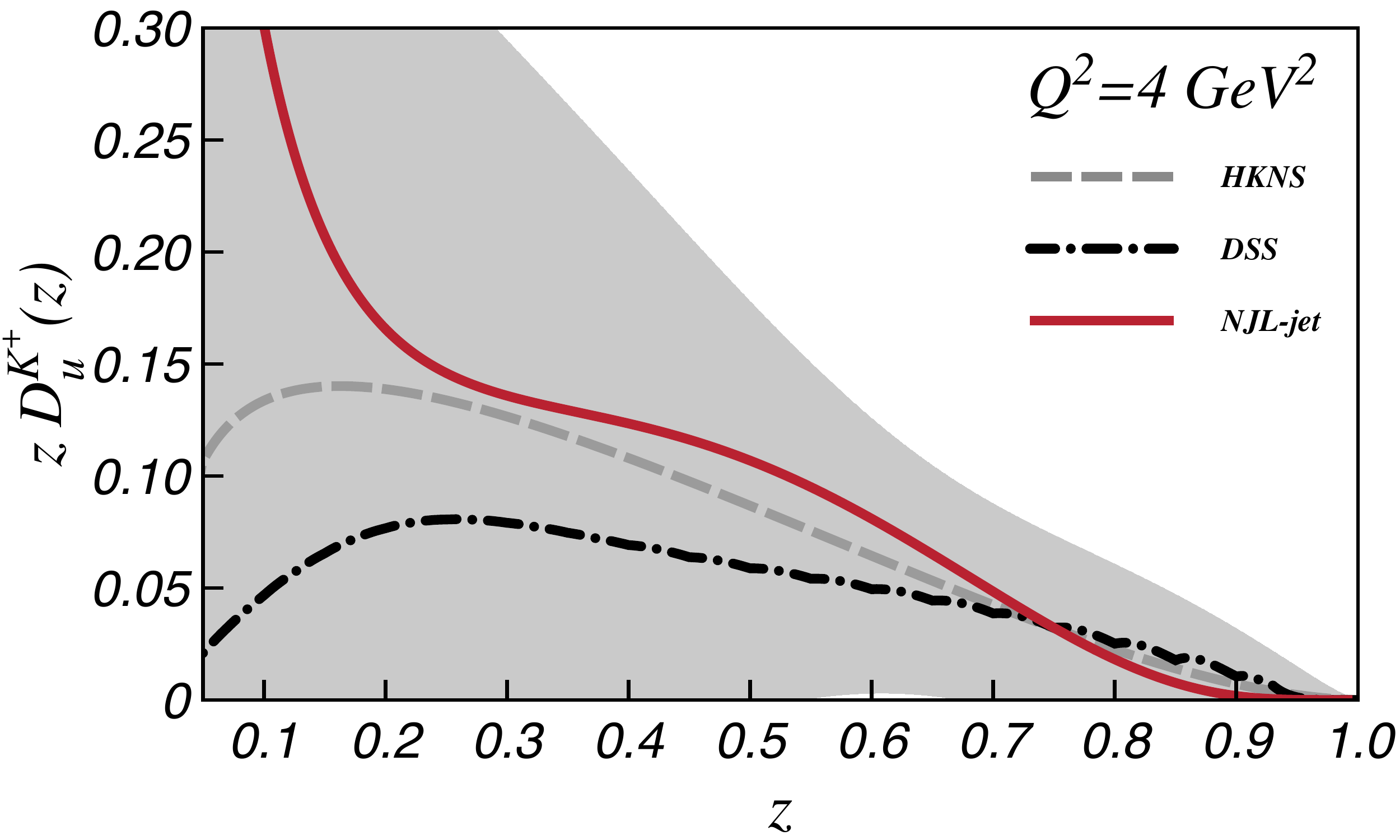}}
\caption{The integrated fragmentation functions for a $u$ quark
to $\pi^+$ (upper) and $K^+$ (lower), calculated using the LB-DIP 
regularization with a light constituent quark mass of $M=0.4\,\mathrm{GeV}$ 
and evolved from the model scale to $Q^2 = 4\,\mathrm{GeV}^2$. The results are compared 
to phenomenological parametrizations of experimental data from 
Ref.~\cite{Hirai:2007cx} (HKNS) and Ref.~\cite{deFlorian:2007aj} (DSS). The 
shaded area represents the uncertainties in the HKNS results.}
\label{PLOT_FRAG_Q_PI_K}
\end{figure}

The plots in Fig.~\ref{PLOT_DRV_LB_DIP} clearly show that the $z$ range of the 
splitting functions calculated using LB-DIP allows for a smooth continuation of the 
corresponding splitting functions calculated using LB regularization to the 
endpoints $z=0$ and $z=1$. The plots in Fig.~\ref{PLOT_FRAG_Q_PI_K} present results
for the fragmentation functions of a $u$ quark to $\pi^+$ and $K^+$ 
using the LB-DIP regularization scheme. We use the QCD evolution code of Ref.~\cite{Botje:2010ay}
at next-to-leading order to evolve our model results from the scale $Q_0^2 = 0.2\,$GeV$^2$
to $Q^2 = 4\,$GeV$^2$.
We find a slightly better description 
of the empirical parametrizations compared to our earlier work~\cite{Matevosyan:2011ey,Matevosyan:2010hh}, 
especially in the region where $z$ is close to $1$. Previously, artifacts of the LB regularization 
did not allow a good description in this domain.

\section{TMD Quark Distributions in the Nucleon}
\label{sec:tmdpdfs}
The TMD quark distributions in the nucleon are again determined by utilizing the NJL
model. The nucleon bound state is described by a relativistic Faddeev equation 
that includes both scalar and axial--vector diquark correlations, where the 
static approximation is used to truncate the quark exchange kernel~\cite{Cloet:2005pp}.
The relevant terms of the NJL interaction Lagrangian are

\begin{align}
&\mathcal{L}_I = G_s \Bigl(\ol{\psi}\,\g_5 C \tau_2 \beta^A\, \ol{\psi}^T\Bigr)
                              \Bigl(\psi^T\,C^{-1}\g_5 \tau_2 \beta_A\, \psi\Bigr) \no \\
&+\,G_a \lf(\ol{\psi}\,\g_\mu C \tau_i\tau_2 \beta^A\, \ol{\psi}^T\rg) 
                              \Bigl(\psi^T\,C^{-1}\g^{\mu} \tau_2\tau_i \beta_A\, \psi\Bigr),
\label{eq:lag}
\end{align}
where $C = i\g_2\g_0$ and $\beta_A = \sqrt{\tfrac{3}{2}}\,\lambda_A~~(A\in2,5,7)$ are the
color $\bar{3}$ matrices~\cite{Cloet:2005pp}. The strength of the scalar and axial--vector diquark
correlations in the nucleon are determined by the couplings $G_s$ and $G_a$, respectively.
To regularize the NJL model for the calculation of the nucleon, we choose the proper-time scheme, 
with an infrared and ultraviolet cutoff, labeled by $\Lambda_{IR}$ and $\Lambda_{UV}$, respectively.
This scheme enables the removal of unphysical thresholds for nucleon decay 
into quarks, and hence simulates an important aspect of confinement~\cite{Ebert:1996vx, Hellstern:1997nv,Bentz:2001vc}.
This simulation of quark confinement has also been shown to provide a natural 
saturation mechanism for nuclear matter in the NJL model~\cite{Bentz:2001vc}.

The proper-time regularization scheme is not used for the fragmentation functions
because the emitted hadrons are not confined. Therefore, the confining nature
of the proper-time regularization is not appropriate in this case. However, for 
consistency between both
regularization schemes we use the same light constituent quark mass and fix the UV
cutoff so as to reproduce the pion decay constant.

\begin{figure}[tbp]
\includegraphics[width=1.0\columnwidth]{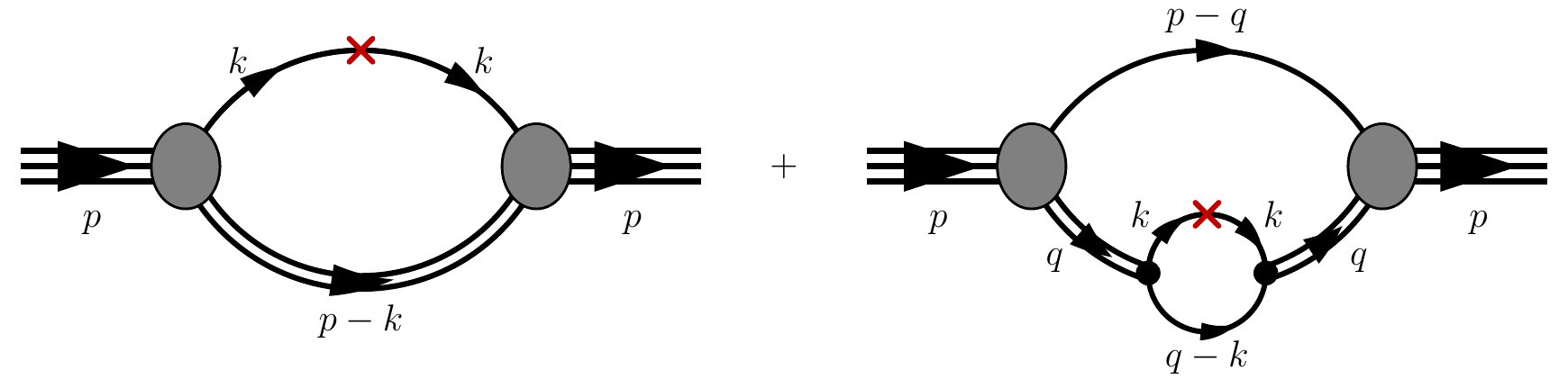}
\caption{Feynman diagrams which give the unpolarized TMD quark distribution functions 
in the nucleon. The single line represents the quark propagator and the double line
the diquark $t$-matrix. The shaded oval denotes the quark-diquark vertex function,
obtained from a relativistic Faddeev equation and the operator insertion has the form 
$\g^+ \delta (x - \tfrac{k^+}{p^+})\tfrac{1}{2}\lf(1\pm\tau_3\rg)$.}
\label{fig:diagrams}
\end{figure}

\begin{figure}[tbp] 
{\subfloat{\centering\includegraphics[width=1.0\columnwidth]{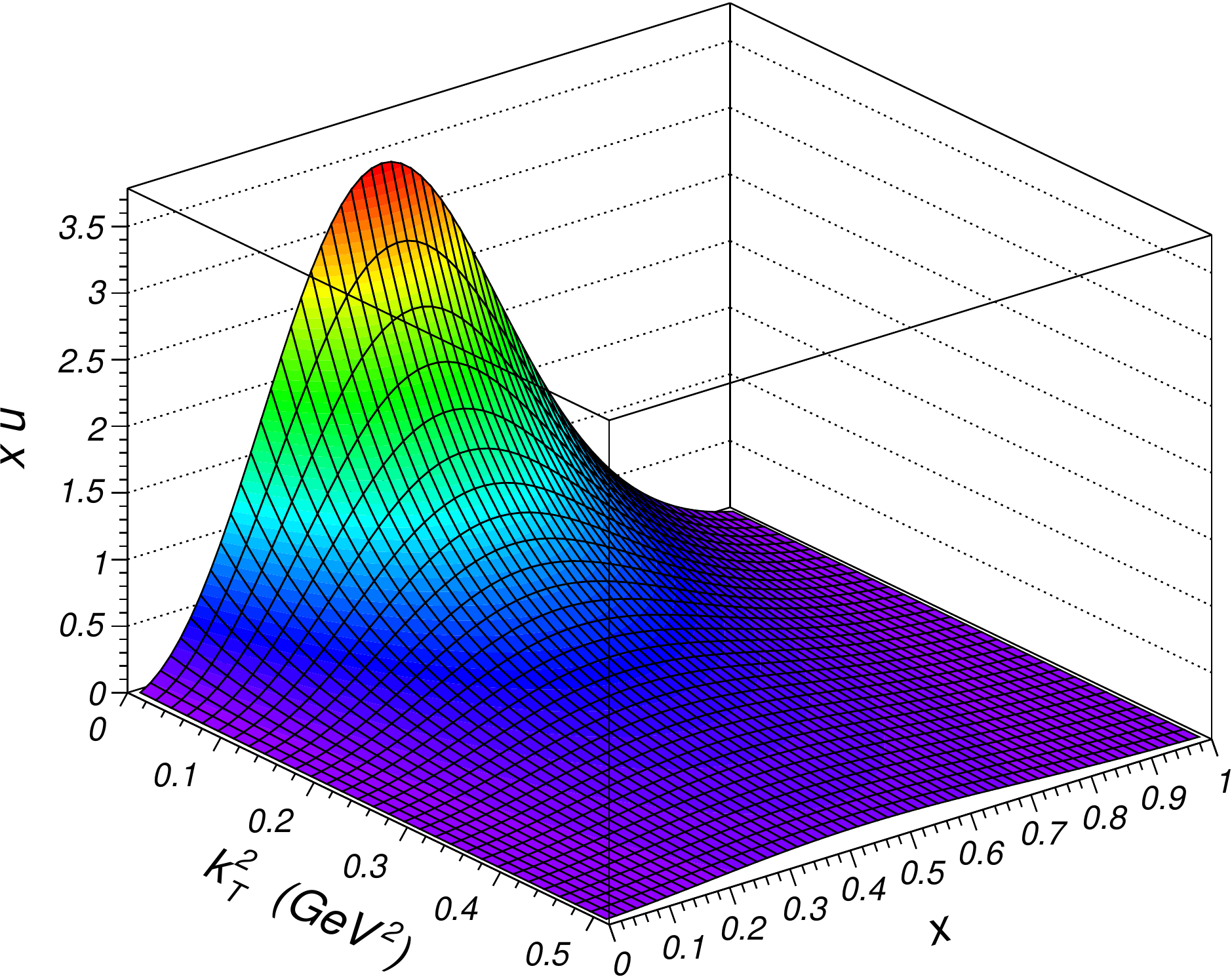}}} \\
{\subfloat{\centering\includegraphics[width=1.0\columnwidth]{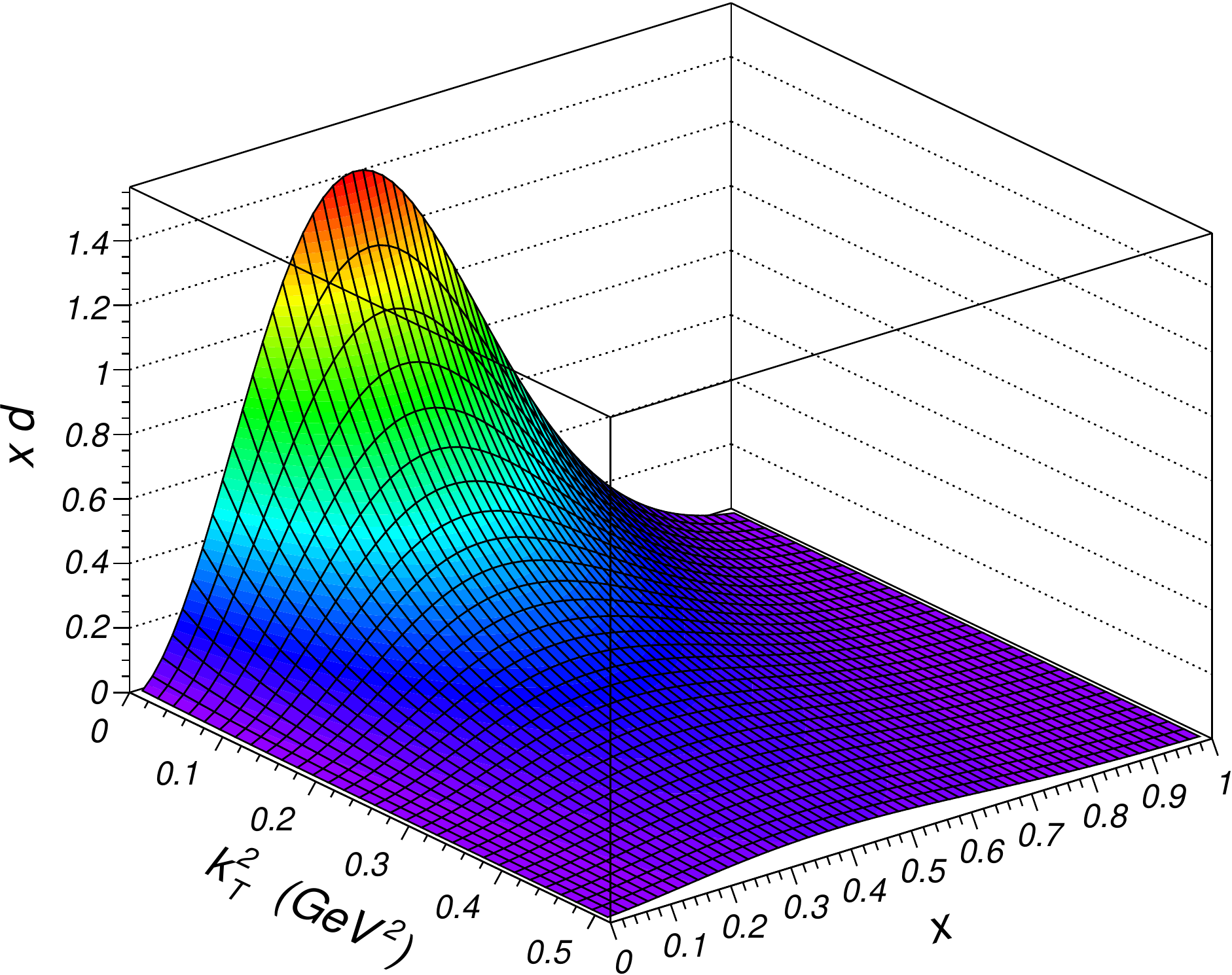}}}
\caption{Results for the $u$ (upper) and $d$ (lower) TMD quark distributions 
in the proton.}
\label{fig:tmds}
\end{figure}

The five parameters of our NJL model for the nucleon are the light constituent quark mass, $M$, the 
regularization parameters $\Lambda_{IR}$ and $\Lambda_{UV}$, and  the couplings $G_s$ 
and $G_a$. These are determined by fixing $M = 0.4\,$GeV, $\Lambda_{IR} = 0.24\,$GeV, and then 
reproducing the nucleon mass, pion decay constant, and the nucleon axial coupling
via Bjorken sum~\cite{Cloet:2006bq,Cloet:2009qs}. Strange quarks are not yet included in our 
model for the nucleon.

The leading-twist spin-independent TMD distribution of the quarks of flavor $q$ in the nucleon is
defined via the correlator~\cite{Bacchetta:2008af,Avakian:2010br}
\begin{multline}
\mathcal{Q}(x,\vect{k_T}) 
= p^+\int \frac{d \xi^- d \vect{\xi}_T}{(2\pi)^3}\ e^{ix\,p^+\,\xi^-}\, e^{-i\,\vect{k_T}\cdot \vect{\xi_T}} \\
\times
\lf\la N,S \lf\vert \bar{\psi}_q(0)\,\g^+\,\mathcal{W}(\xi)\,\psi_q(\xi^-,\xi_T)  \rg\vert N,S \rg\ra \Bigr\vert_{\xi^+=0},
\label{eq:tmd1}
\end{multline}
where $\mathcal{W}(\xi)$ is a gauge link connecting the two quark fields, which are labeled by $\psi_q$.
In QCD this gauge link is nontrivial for $\vect{\xi_T} \neq 0$, however at the level of approximation
that we are working at, this gauge link equals unity in the NJL model.
Our states are normalized using the noncovariant light-cone normalization, namely
\begin{align}
\sideset{}{_q}\sum \lf\la N,S \lf\vert \bar{\psi}_q(0)\,\g^+\,\psi_q(0)  \rg\vert N,S \rg\ra = 3.
\end{align}
Equation~\eqref{eq:tmd1} can be expressed in terms of two TMD quark distribution functions,
namely,
\begin{align}
\mathcal{Q}(x,\vect{k_T})  &= q(x,k_T^2) - \frac{\ve^{-+ij}\,k_T^i\,S_T^j}{M}\, q_{1T}^\perp(x,k_T^2),
\end{align}
where the first TMD PDF integrated over $\vect{k_T}$ gives the familiar unpolarized 
quark distribution function and the second TMD PDF, known as the Sivers function~\cite{Sivers:1989cc,Bacchetta:2003rz}, 
is time-reversal odd and is zero at the level of approximation included in this work.

To determine the TMD quark distributions in this model, it is convenient to express 
them in the form \cite{Jaffe:1985je,Barone:2001sp}
\begin{align}
q(x,\,k_T^2) = -i\int \frac{d k^+ dk^-}{(2\pi)^4}
\delta\!\lf(x - \frac{k^+}{p^+}\rg) \text{Tr}\lf[\g^+\,M_q(p,k)\rg],
\label{eqn:def2}
\end{align}
where $M_q(p,k)$ is the quark two-point function in the bound nucleon. 
Therefore, within any model that describes the nucleon as a bound state of quarks,
the quark distribution functions can be associated with a straightforward Feynman
diagram calculation. 

The Feynman diagrams considered here are given in Fig.~\ref{fig:diagrams}, where
the first diagram represents the so--called quark diagram and the second the diquark 
diagram. The single line in each diagram represents a quark propagator which is the
solution to the gap equation and the double line is the diquark $t$-matrix obtained from
the Bethe--Salpeter equation. The vertex functions represent the solution to
the nucleon Faddeev equation. The resulting distributions have no support for negative $x$
and therefore this is essentially a valence quark picture.
By separating the isospin factors, the spin-independent
$u$ and $d$ TMD quark distributions in the proton can be expressed as 
\begin{align}
\label{eq:uptmd}
u(x,k_T^2) &= f^s_{q/N}(x,k_T^2)  + \tfrac{1}{3}\,f^a_{q/N}(x,k_T^2)  \no \\
&\hs{0mm} + \tfrac{1}{2}\,f^s_{q(D)/N}(x,k_T^2) + \tfrac{5}{6}\,f^a_{q(D)/N}(x,k_T^2),\\[1.0ex]
\label{eq:downtmd}
d(x,k_T^2) &= \tfrac{2}{3}\,f^a_{q/N}(x,k_T^2)  \no \\
&\hs{0mm} + \tfrac{1}{2}\,f^s_{q(D)/N}(x,k_T^2) + \tfrac{1}{6}\,f^a_{q(D)/N}(x,k_T^2).
\end{align}
The superscripts $s$ and $a$ refer to the scalar and axial-vector terms, respectively, 
the subscript $q/N$ implies a quark diagram and $q(D)/N$ a diquark diagram. Explicit expressions 
for the functions in Eqs.~\eqref{eq:uptmd} and \eqref{eq:downtmd} are given in the
Appendix.

\begin{figure}[tbp]
\includegraphics[width=1.0\columnwidth]{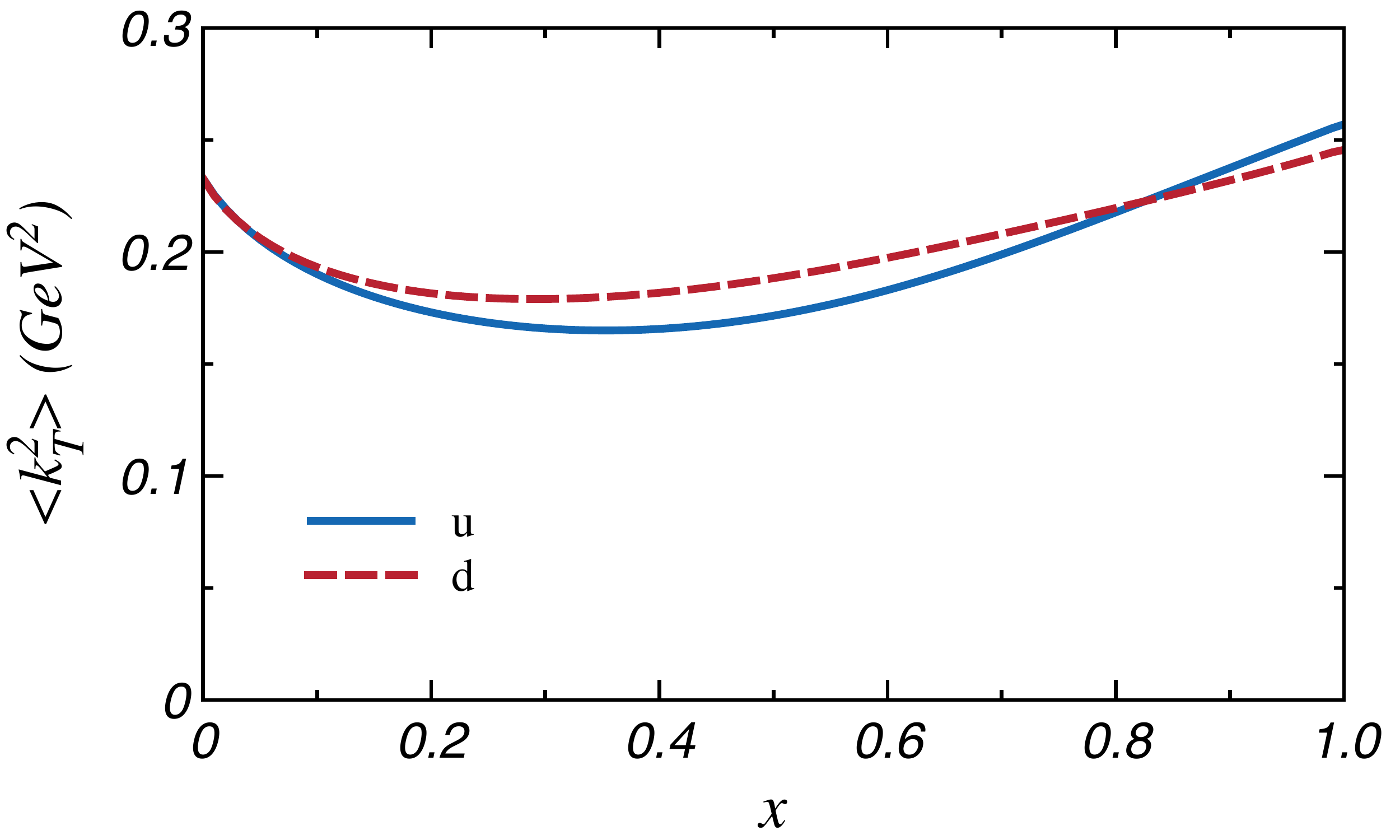}
\caption{The Bjorken $x$ dependence of $\lf<k_T^2\rg>$. Diquark correlations
in the nucleon give rise to the quark flavor dependence.}
\label{fig:kT2vx}
\end{figure}

\begin{figure}[tbp]
\subfloat{\centering\includegraphics[width=1.0\columnwidth]{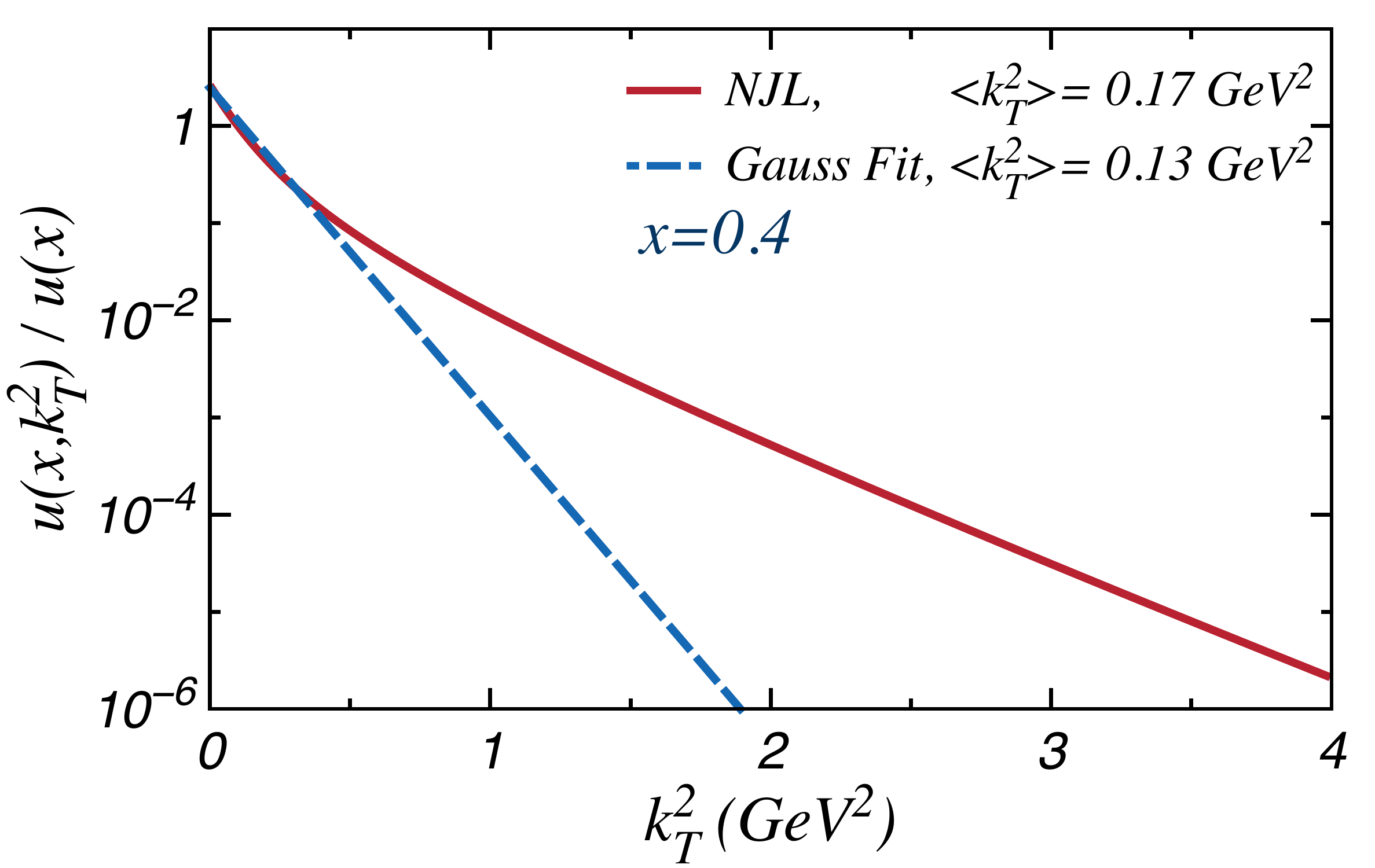}}\\
\subfloat{\centering\includegraphics[width=1.0\columnwidth]{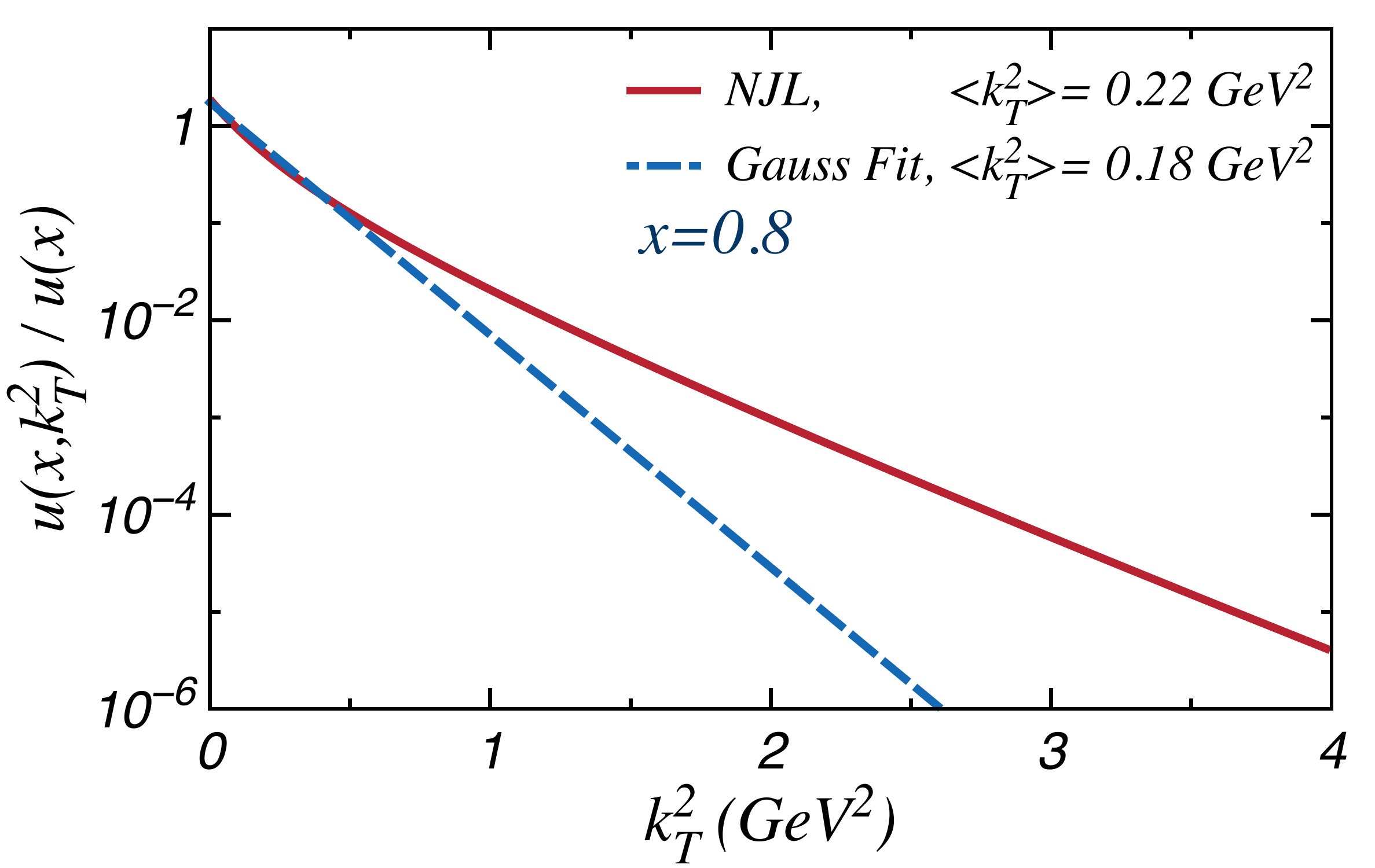}}
\caption{Results for the TMD $u$-quark distribution in the proton for fixed $x$ slices,
where the upper plot has $x=0.4$ and the lower plot is for $x=0.8$.
Also plotted are individual fits to the TMD quark distribution using the Gaussian ansatz of 
Eq.\eqref{eq:tmdgaussian} for each $x$, with $\langle k_T^2\rangle$ the single fit parameter 
in each case.}
\label{fig:tmdsfixedx}
\end{figure}

Results for the $u$- and $d$-quark TMD quark distributions functions in the proton
are illustrated in Fig.~\ref{fig:tmds}. The $Q^2$ scale to which these results 
correspond is not determined by the model. Previously for the familiar spin-independent
PDFs we fitted the valence $u$-quark distribution in the proton to the empirical result
at some large $Q^2$ scale, this gives a model scale of 
$Q_0^2 = 0.16\,$GeV$^2$~\cite{Cloet:2005pp,Cloet:2006bq,Cloet:2007em} in the proper-time
regularization scheme. Rigorous comparison with the experimental data requires QCD evolution of the model TMD PDFs, which is left for future work. Here, we just show the results as they emerge from our model, the
exact scale of which is not so important for this purpose. When QCD evolution is included, both the TMD PDF and TMD fragmentation function model scales must be equal when determining observables like SIDIS cross--sections.
The integral of these TMD PDF results over $\vect{k_T}$ gives the familiar spin-independent 
quark distributions functions, which satisfy the baryon number and momentum sum rules. 
The Bjorken $x$ and $k_T^2$ dependence in these expressions is not separable, and 
therefore the Gaussian ansatz for the TMD quark distributions, namely, that they can 
be written in the form
\begin{align}
q(x,\,k_T^2) = q(x)\ \frac{e^{-k_T^2/\lf<k_T^2\rg>}}{\pi\,\lf<k_T^2\rg>},
\label{eq:tmdgaussian}
\end{align}
is not possible for our TMD PDF results. The Bjorken $x$ dependence of 
$\lf<k_T^2\rg>$  for our proton TMD quark distribution results is illustrated
in Fig.~\ref{fig:kT2vx}, where
\begin{align}
\lf<k_T^2\rg>(x) \equiv \frac{\int d^2 \vect{k_T}\ k_T^2\, q(x,\,k_T^2)}{\int d^2 \vect{k_T}\ q(x,\,k_T^2)}.
\label{eq:tmdkt2}
\end{align}
If the $x$ and $k_T^2$ dependence of our TMD quark distributions were separable then
the curves in Fig.~\ref{fig:kT2vx} would be constants, however we find that $\lf<k_T^2\rg>$
has about a $20\%$ variation over the domain of Bjorken $x$. We also find that the
$x$ dependence of $\lf<k_T^2\rg>$ for the $u$ and $d$ quarks differs somewhat, with the 
$d$ quarks having slightly larger $\lf<k_T^2\rg>$ for the majority of Bjorken $x$. 

\begin{figure}[tbp] 
\subfloat{\centering\includegraphics[width=1.0\columnwidth]{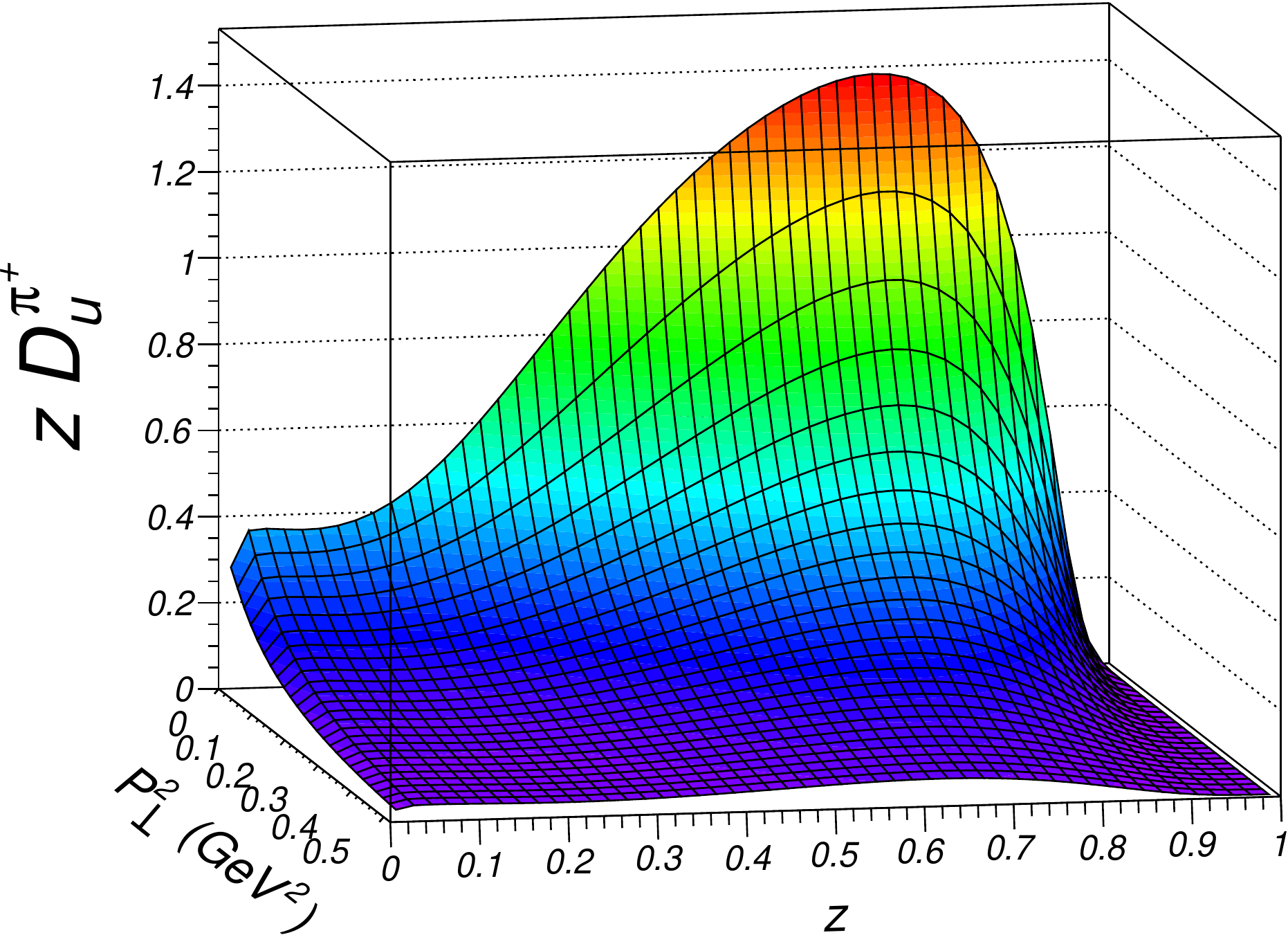}} \\
\subfloat{\centering\includegraphics[width=1.0\columnwidth]{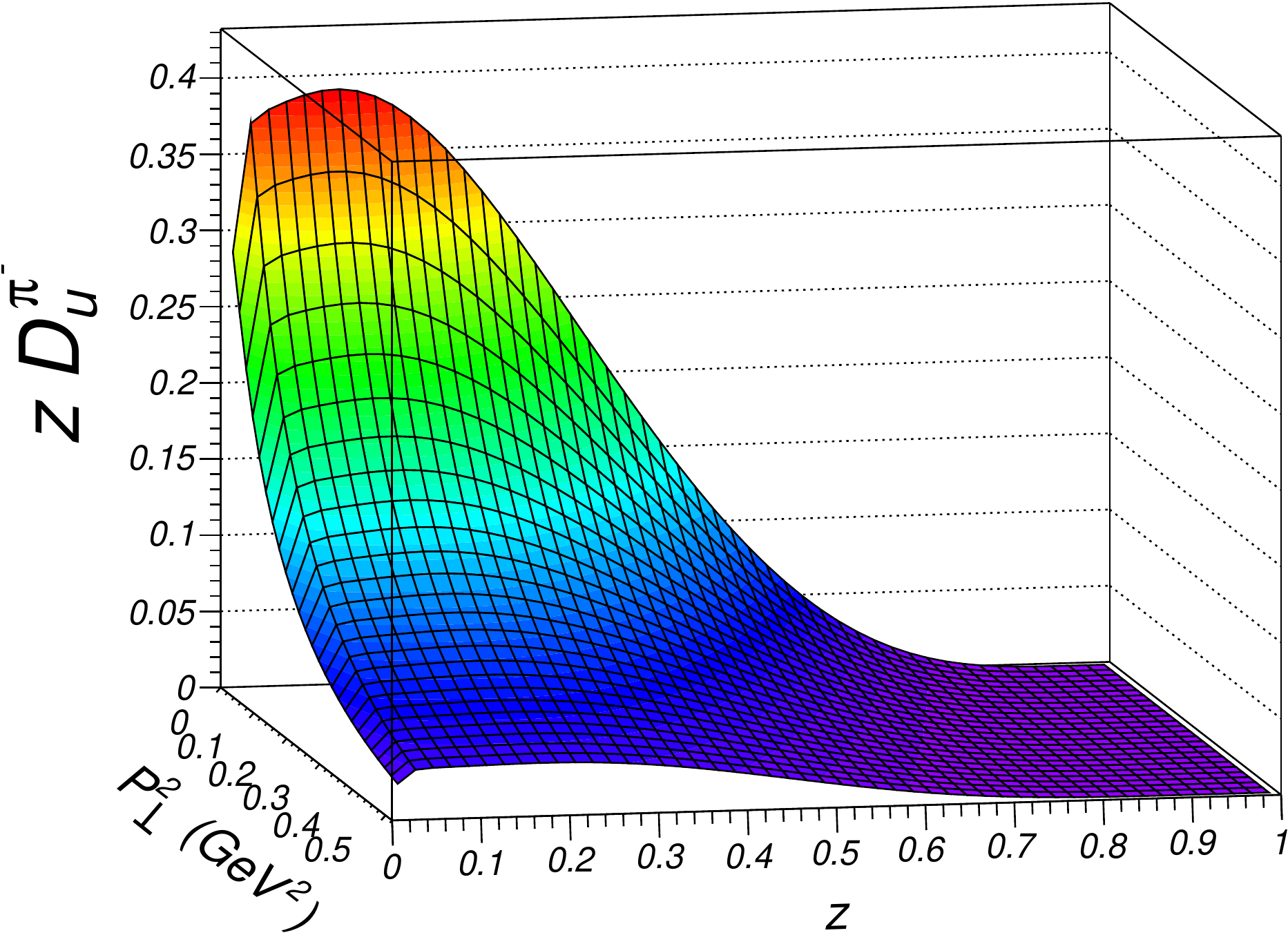}}
\caption{TMD fragmentation functions for a $u$ quark to $\pi^+$ and $\pi^-$. 
The upper figure illustrates the favored case, which peaks at relatively large $z$,
while the unfavored case, shown in the lower figure, peaks at much smaller $z$.}
\label{PLOT_JET_TMD_PI}
\end{figure}

\begin{figure}[tbp] 
\subfloat{\centering\includegraphics[width=1.0\columnwidth]{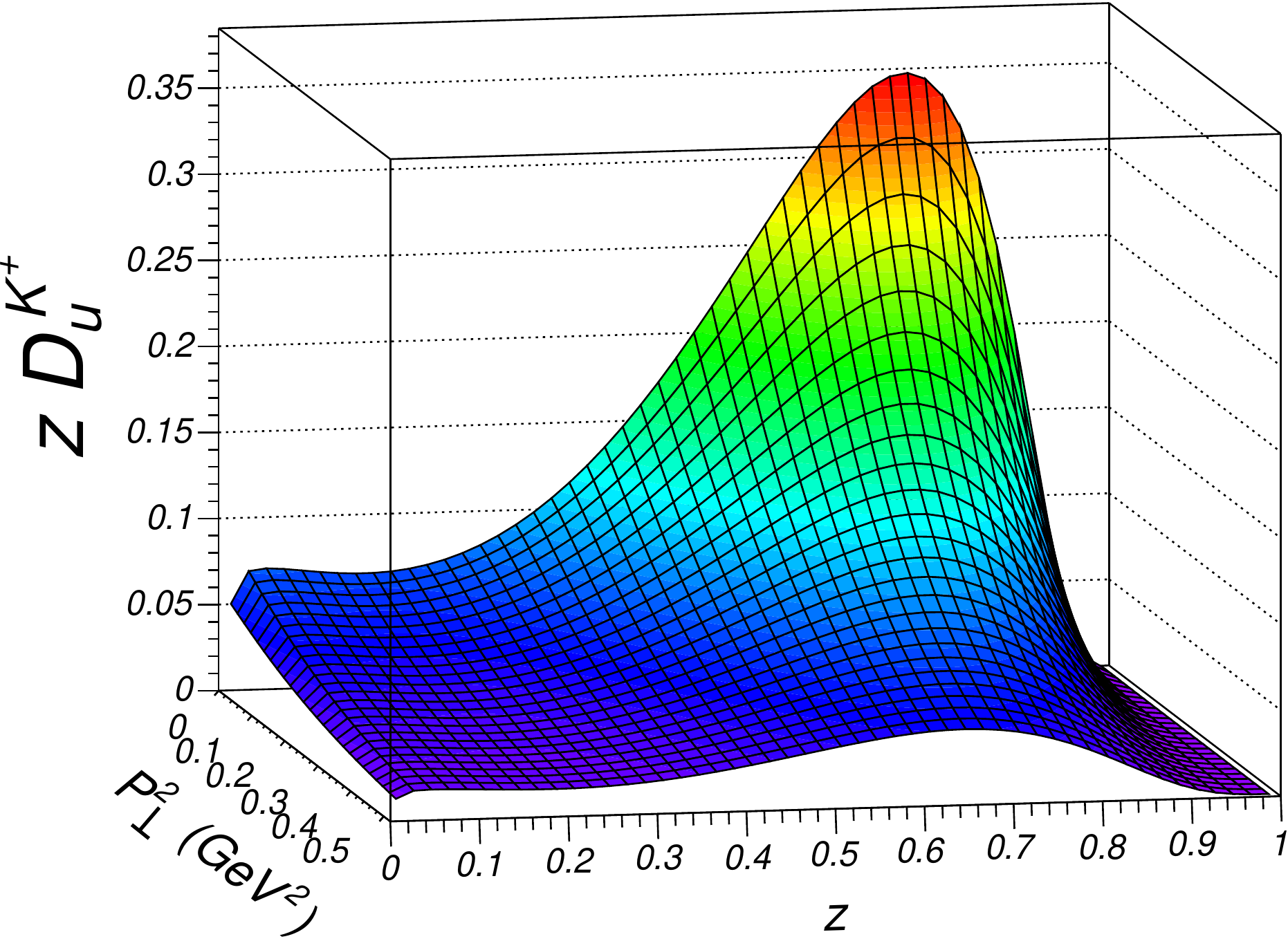}} \\
\subfloat{\centering\includegraphics[width=1.0\columnwidth]{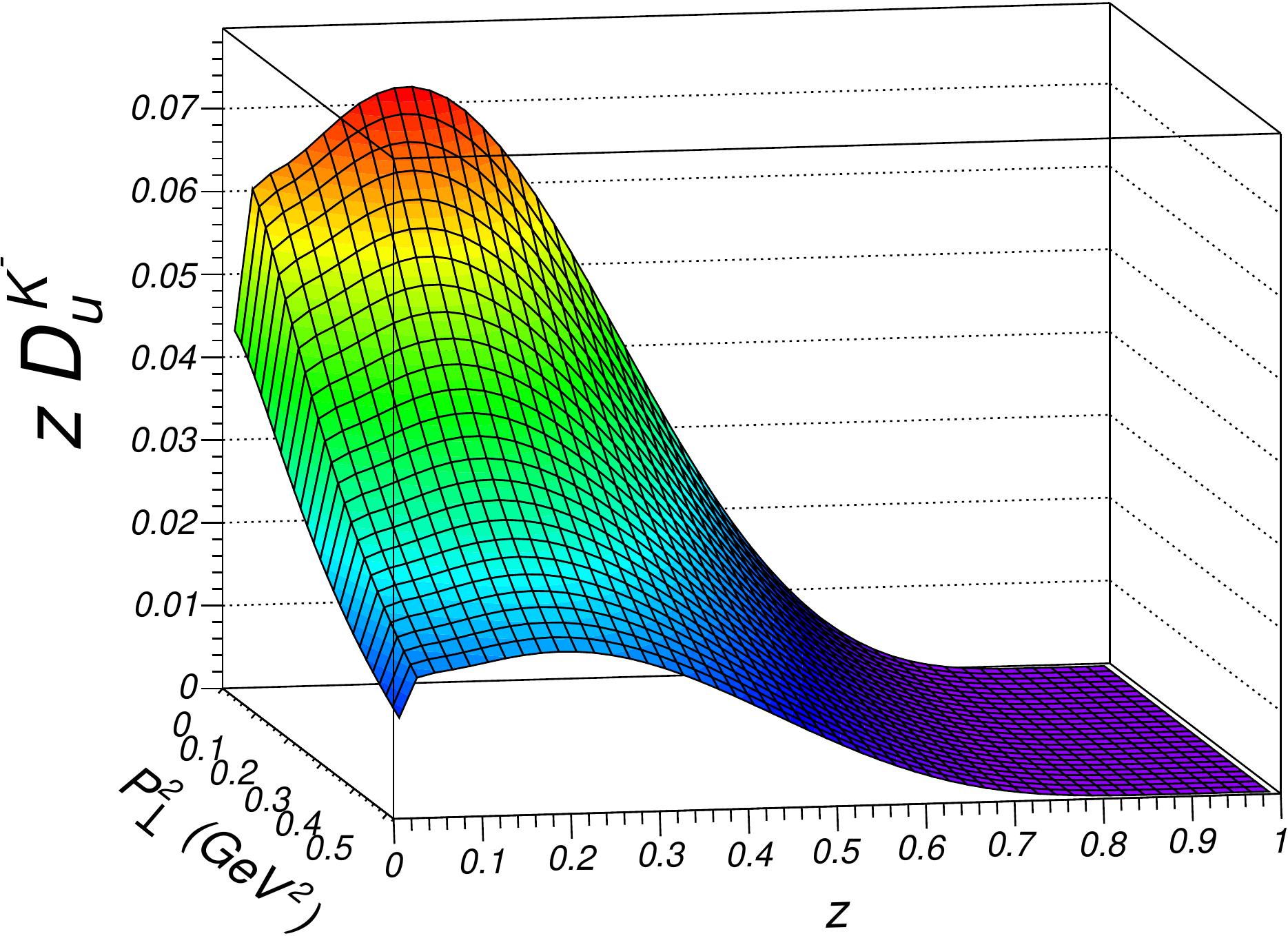}}
\caption{TMD fragmentation functions for a $u$ quark to $K^+$ and $K^-$. 
The upper figure illustrates the favored case, which peaks at relatively large $z$,
while the unfavored case, shown in the lower figure, peaks at much smaller $z$.}
\label{PLOT_JET_TMD_K}
\end{figure}

Figure~\ref{fig:tmdsfixedx} 
illustrates our TMD quark distribution results at particular $x$ values and compares
them to a Gaussian ansatz fit for the same $x$ slice. 
The Gaussian ansatz results are obtained by a least squares fit of the TMD factor
in Eq.~\eqref{eq:tmdgaussian} to our ratios $q(x,k_T^2)/q(x)$ calculated in the NJL model,
using $\langle k_T^2 \rangle$ of Eq.~\eqref{eq:tmdgaussian} as the only fit parameter for 
each value of $x$. The fitted value of this parameter is approximately 20\% smaller than the
value of $\langle k_T^2 \rangle$ calculated with our model distribution functions. In the
least squares fit, we included values of $k_T^2$ up to $4\,$GeV$^2$ and the curves in
Fig.~\ref{fig:tmdsfixedx} indicate that such a fit to a single Gaussian is reasonable 
only for a limited $k_T^2$ region, for a single value of $x$.

\section{TMD Fragmentation Function Results}
\label{sec:results}

In this section, we present NJL-jet model results for the TMD fragmentation functions. The
number of emitted hadrons in the decay chain is set to $N_{Links}=6$, which is sufficient
to accurately obtain the pion and kaon fragmentation functions in the domain $z \gtrsim 0.02$. 
We solve for the fragmentation of $u$, $d$, and $s$ quarks to pions and kaons, 
utilizing Monte Carlo simulations and the expression in Eq.~(\ref{EQ_FRAG_MC_TMD}), similar 
to our previous calculations of the integrated fragmentation functions detailed 
in Ref.~\cite{Matevosyan:2011ey}. The computational challenge for the Monte Carlo simulations 
is to obtain sufficient statistics and this becomes significantly more difficult when we include 
the transverse momentum dependence, because now the number of bins becomes quadratic in the size of the discrete 
bin size (taken to be $1/500$ both for $z$ and transverse momentum, in the 
corresponding units). Furthermore, the extent of the bins in the transverse momentum 
direction was extended to $6~\mathrm{GeV}^2$, in order to avoid any notable numerical 
artifacts arising from the limited range of transverse momentum. To overcome the 
numerical challenge, our software platform was developed to allow for parallel 
generation of the Monte Carlo quark decay cascades, with different seeds for their random 
number generators. The results were later combined to produce the high statistics 
solutions. The computations were facilitated on the small computer cluster at 
the Special Research Centre for the Subatomic Structure of Matter (CSSM)
that consists of 11 machines with Intel Core i7 920 quad core CPUs  running on 
the Linux Fedora Core 11 operating system and GCC 4.4. A typical calculation 
of fragmentation for a given quark type takes about $12$ hours with $44$ parallel 
processors.

Results for the TMD favored and unfavored fragmentation functions for a $u$ quark to 
$\pi$ and $K$ mesons are illustrated in Figs.~\ref{PLOT_JET_TMD_PI} and \ref{PLOT_JET_TMD_K}.
In each case, the favored TMD fragmentation functions have more support at large $z$, while
the unfavored results are peaked at smaller $z$. It is also evident that the kaon
fragmentation functions fall off more slowly in $P_\perp^2$ than the 
corresponding pion fragmentation functions. The drop in each of the fragmentation functions
for $z \lesssim 0.02$ is a consequence of choosing $N_{Links} = 6$, which means that in
the Monte Carlo simulation there is a vanishingly small probability of emitting hadrons 
with $z < 0.02$.

\begin{figure}[tbp]
\includegraphics[width=1.0\columnwidth]{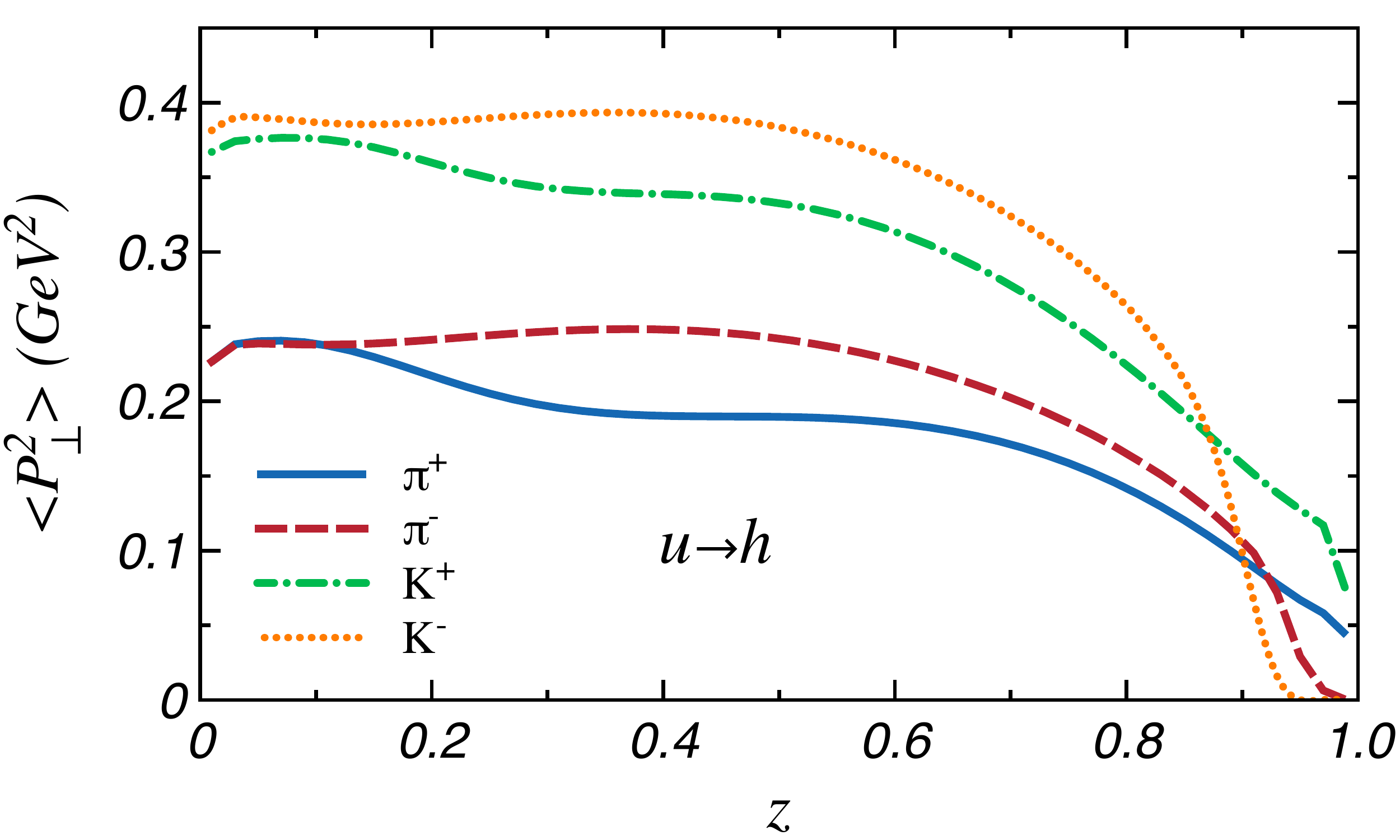}
\caption{The averaged transverse momentum of $\pi$ and $K$ mesons emitted by a 
$u$ quark.}
\label{PLOT_JET_TMD_PP_PI_K}
\end{figure}

The Gaussian ansatz is widely used to describe the traverse momentum dependence of both quark distribution 
and fragmentation functions.
In particular, the TMD fragmentation function of a quark $q$ emitting a hadron $h$ is often modeled by
\begin{equation}
\label{EQ_FF_GAUSS}
D_q^h(z,P_\perp^2)= D_q^h(z) \frac{e^{-P_\perp^2/\langle P_\perp^2 \rangle}}{\pi \langle P_\perp^2 \rangle},
\end{equation}
where $D_q^h(z)$ is the corresponding integrated fragmentation function and $\langle P_\perp^2 \rangle$ 
is the average transverse momentum of the produced hadron $h$, defined by
\begin{align}
\langle P_\perp^2 \rangle(z)  \equiv 
\frac{\int d^2 \vect{P_\perp}\ P_\perp^2\, D_q^h(z, P_\perp^2)}{\int d^2 \vect{P_\perp}\ D_q^h(z, P_\perp^2)}.
\label{eq:p_perp2}
\end{align}
In analyses that assume a Gaussian ansatz for the TMD fragmentation functions,
it is usual to assume that $\langle P_\perp^2 \rangle$ does not depend on $z$, the type of hadron, $h$, 
or the quark flavor, $q$.
These assumptions will be tested against the  NJL-jet TMD fragmentation functions. 

\begin{figure}[tbp]
\centering 
\subfloat{\centering\includegraphics[width=1.0\columnwidth]{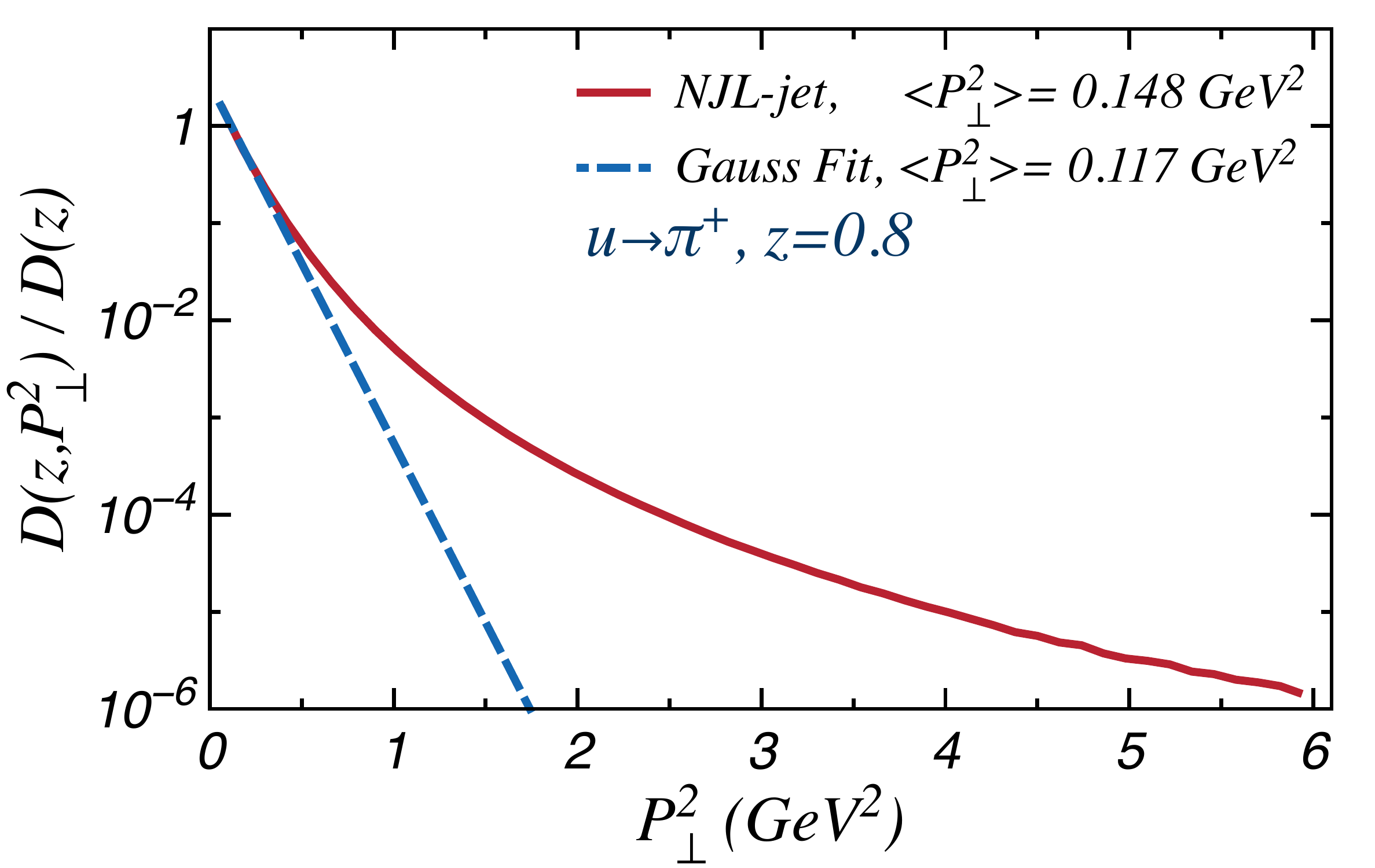}}\\
\subfloat{\centering\includegraphics[width=1.0\columnwidth]{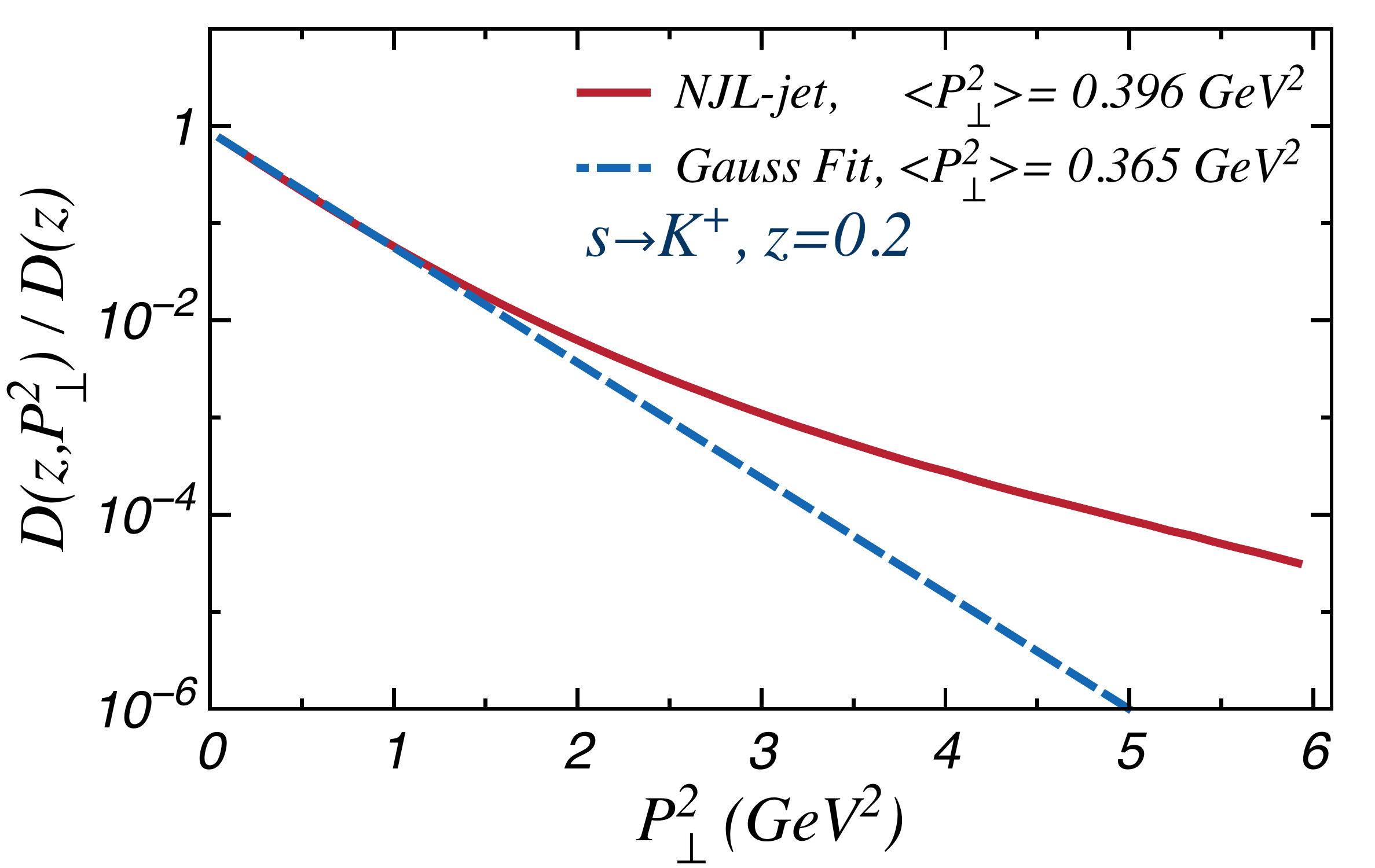}}
\caption{Normalized TMD fragmentation for the favored $u\to\pi^+$ process for $z=0.8$ (upper)
and the unfavored $s\to K^+$ process for $z=0.2$ (lower). Also depicted are fits to the fragmentation 
functions using the Gaussian ansatz of Eq.~\eqref{EQ_FF_GAUSS}, with $\langle P_\perp^2 \rangle$ 
as the single fitting parameter.}
\label{PLOT_PPSQ_PROJ}
\end{figure}

The results in Fig.~\ref{PLOT_JET_TMD_PP_PI_K} depict the average transverse momenta of $\pi$ and $K$ mesons 
produced by a $u$-quark. These plots 
show that the average transverse momenta of the hadrons are relatively flat versus $z$ in the region $0.3 < z < 0.6$, 
however they have a significant dependence on the type of the hadron. We find that the average transverse momentum of the 
kaons is significantly larger than that of the pions.

The curves in Fig.~\ref{PLOT_PPSQ_PROJ} depict the TMD fragmentation of a favored $u\to\pi^+$ 
process for $z=0.8$ and an unfavored $s\to K^+$ process for $z=0.2$. Also presented are least squares fits 
to the fragmentation functions for particular $z$ slices using the Gaussian ansatz of Eq.~\eqref{EQ_FF_GAUSS}, 
with $\langle P_\perp^2 \rangle$ the single fitting parameter for each $z$. 
The plots in Fig.~\ref{PLOT_PPSQ_PROJ} indicate that such a fit to a single Gaussian is reasonable only 
for a limited $P_\perp^2$ region. Also, because $\langle P_\perp^2 \rangle$ has a significant $z$ 
dependence, the Gaussian ansatz for the entire TMD fragmentation function offers at best a crude 
approximation to the full results.
The corresponding average transverse momenta obtained from the Gaussian fits are smaller than 
those obtained directly using the relation in Eq.~\eqref{eq:p_perp2}.

\section{Average Transverse Momenta in SIDIS}
\label{sec:sidis}

\begin{figure}[tbp]
\includegraphics[width=1.0\columnwidth]{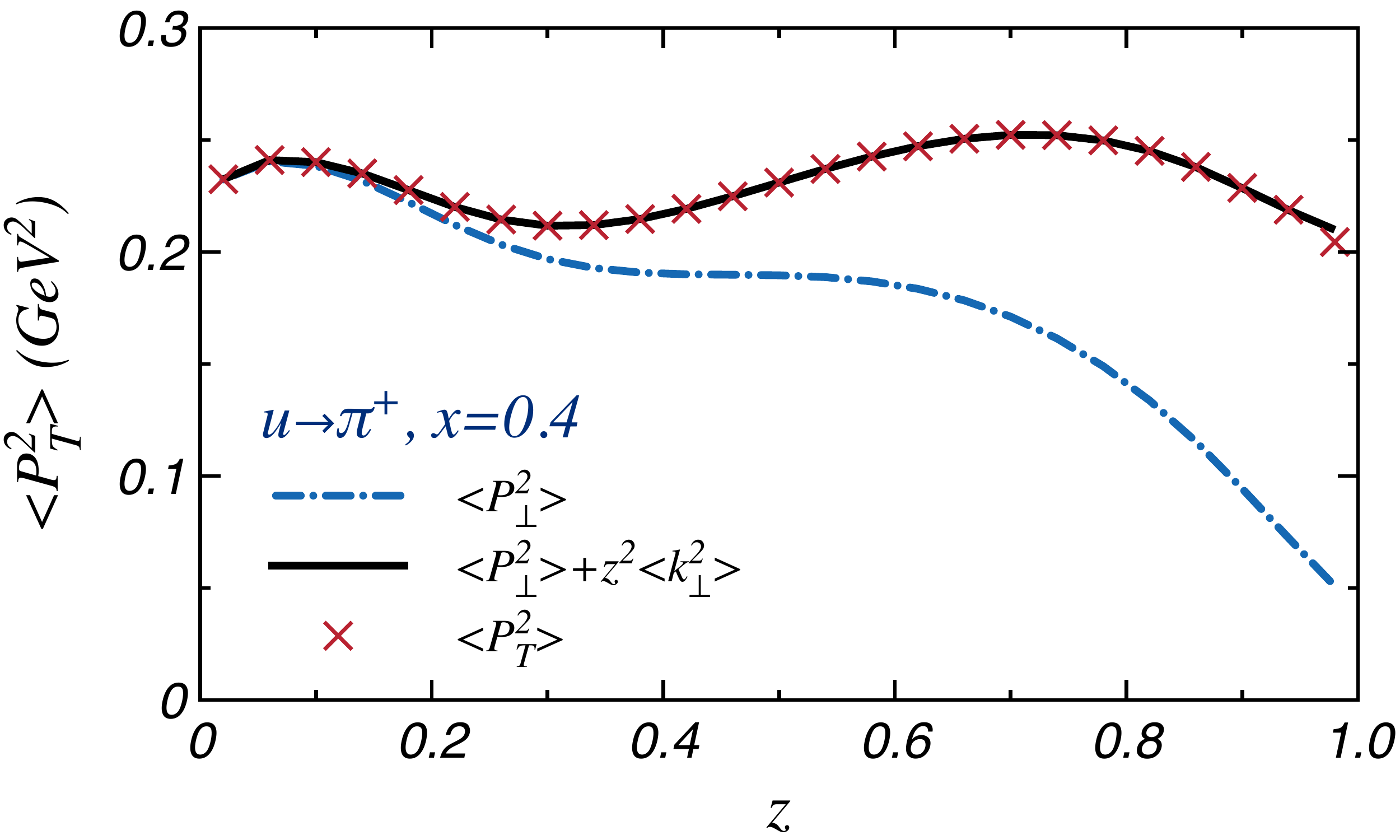}
\caption{The averaged transverse momentum of $\pi^+$ mesons in SIDIS produced 
on a $u$ quark in a proton with light-cone momentum fraction $x=0.4$.}
\label{PLOT_JET_TMD_PT_PIPL}
\end{figure}
\begin{figure}[tbp]
\includegraphics[width=1.0\columnwidth]{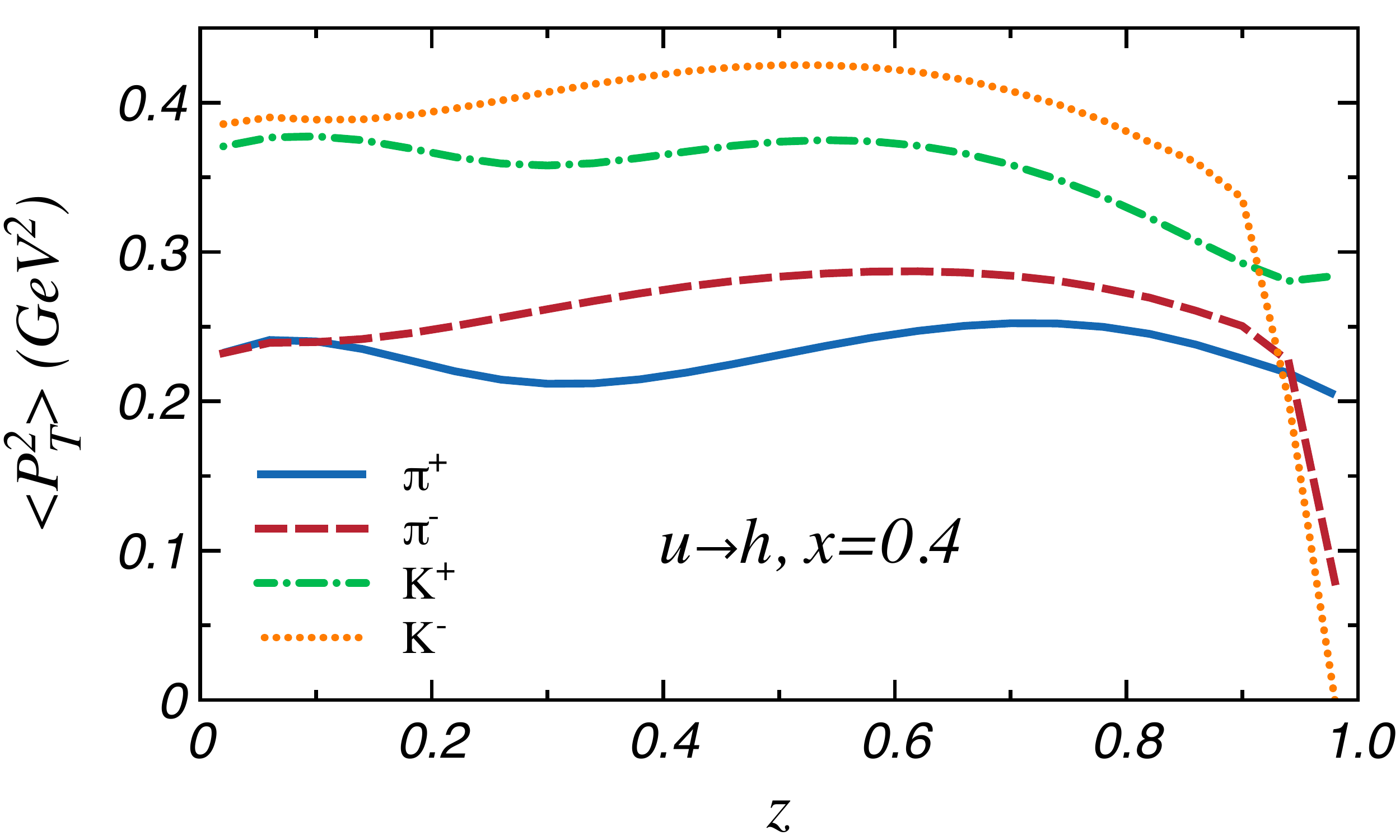}
\caption{The averaged transverse momentum of $\pi$ and $K$ mesons in SIDIS on a 
$u$ quark in the proton with light-cone momentum fraction $x=0.4$. 
The unfavored fragmentation functions rapidly approach zero for large $z$ and this
causes the dramatic changes in $\langle P_T^2\rangle$ at large $z$ illustrated in 
this figure.}
\label{PLOT_JET_TMD_PT_PI_K}
\end{figure}

For the SIDIS process we have created a Monte Carlo event generator that can calculate
the physical cross-section. In future work this will enable us to analyze the relative importance of the different aspects of the process and the implications of the constraints set in individual experiments.
We use it to determine the average transverse 
momentum of the produced hadrons (at the model scale) observed in a SIDIS experiment, namely $\langle P_T^2 \rangle$, which is 
defined as
\begin{align}
\langle P_T^2 \rangle(x,z) \equiv \frac{\int d^2 \vect{P_T}\ P_T^2\, \widetilde{D}_q^h(x,z, P_T^2)}
                                  {\int d^2 \vect{P_\perp}\ \widetilde{D}_q^h(x,z, P_T^2)}.
\end{align}
The function $\widetilde{D}_q^h(x,z, P_T^2)$ is defined in Eq.~\eqref{EQ_SIDIS_X_SEC}.
The crosses in Fig.~\ref{PLOT_JET_TMD_PT_PIPL} represent results for $\langle P_T^2 \rangle$
acquired by $\pi^+$ mesons in a SIDIS hadronization process, where the virtual photon strikes
a valence $u$ quark in a proton carrying a light-cone momentum fraction of $x=0.4$.
We also plot as the dash-dotted line $\langle P_\perp^2 \rangle$, which is the average transverse 
momentum that the $\pi^+$ mesons acquire in the quark fragmentation process.
Recall, that the transverse momentum
$\vect{P_\perp}$ is defined relative to the direction of the original fragmenting quark,
while $\vect{P_T}$ is relative to the direction of the photon momentum, these transverse
momenta are related by Eq.~\eqref{EQ_PT_PP}. For the factorization of the SIDIS cross--section
given in Eq.~\eqref{EQ_SIDIS_X_SEC}, it can be shown that $\langle P_T^2 \rangle$ is given
by 
\begin{equation}
\label{EQ_PT_PP_AV}
\langle P_T^2 \rangle(x,z) = \langle P_\perp^2 \rangle(z) + z^2 \langle k_T^2 \rangle(x).
\end{equation}
As an additional check on the Monte Carlo calculation, in Fig.~\ref{PLOT_JET_TMD_PT_PIPL} we plot
the result obtained from Eq.~\eqref{EQ_PT_PP_AV} as the solid line and find that it agrees perfectly
with that obtained from the Monte Carlo event generator for the SIDIS cross--section.
We also find that both $\langle P_\perp^2 \rangle$ and $\langle P_T^2 \rangle$ illustrated in
Fig.~\ref{PLOT_JET_TMD_PT_PIPL} have a sizable $z$ dependence.

Illustrated in Fig.~\ref{PLOT_JET_TMD_PT_PI_K} are results for the average transverse 
momentum acquired by $\pi$ and $K$ mesons in the hadronization process in SIDIS, where
the struck quark is a $u$ quark in a proton with light-cone momentum fraction $x=0.4$. 
The rapid approach to zero for the unfavored fragmentation functions in Fig.~\ref{PLOT_JET_TMD_PT_PI_K} 
is a consequence of the large $z$ behavior of the unfavored $\langle P_\perp^2\rangle$
illustrated in Fig.~\ref{PLOT_JET_TMD_PP_PI_K}, which also rapidly approach zero.
The HERMES experimental results for $\langle P_T^2 \rangle$ measured in SIDIS on 
a deuterium target~\cite{Airapetian:2009jy}, are of comparable size to our results
shown in Fig.~\ref{PLOT_JET_TMD_PT_PI_K}. 
We do not plot these HERMES results because the kinematic range is too different
for a quantitative comparison.
The average transverse momentum of the kaons is larger than that of the 
pions at the low $Q^2$ scale of the  model. Our model includes only the valence quarks 
in the proton, which should be the dominant component at $x=0.4$. 
 
\section{Conclusions and Outlook}
\label{sec:conclusions}

In this work we extended the NJL-jet model to include the transverse momentum
dependence in the quark hadronization process. This was achieved using TMD elementary fragmentation
functions and by keeping track of the quark's recoil transverse momentum in the
hadron emission cascade. We modified the LB regularization scheme
to remove artifacts that limit the $z$ range of the splitting
functions, and this in turn improved our description of the integrated fragmentation
functions. The TMD fragmentation functions for $u$, $d$, and $s$
quarks to pions and kaons were determined using a Monte Carlo approach. The average
$P_\perp^2$ of the produced kaons was found to be significantly larger than that of the pions
and in both cases $\la P_\perp^2 \ra$ had a sizable $z$ dependence.
The high statistical precision needed for these calculations was achieved through 
parallel computing on the small computer farm at CSSM. 

The TMD quark distribution functions in the proton were also determined using the NJL model.
In this case, we used the proper-time regularization scheme, because this
method simulates important aspects of confinement. Our TMD PDF results when integrated
over $\vect{k_T}$ give our earlier results for the familiar spin-independent quark
distribution functions~\cite{Cloet:2005pp}, whose moments satisfy the baryon number 
and momentum sum rules.
We found that the average $k_T^2$ of the quarks in the nucleon have a significant 
$x$ dependence and therefore the familiar Gaussian ansatz for the TMD PDFs produces 
only a crude approximation to our full TMD PDF results.

Finally, using the TMD quark distribution functions for the nucleon and 
the results for the TMD fragmentation functions, we constructed 
a Monte Carlo event generator for the SIDIS process. Using this Monte Carlo event 
generator, we determined the average transverse momentum of the hadrons, $\la P_T^2\ra$,  
produced in SIDIS. These results are of a similar magnitude to those extracted from 
experiment, even at our relatively low model scale. As a cross check for this
SIDIS Monte Carlo event generator, we compared our results for $\la P_T^2\ra$ with
those obtained using Eq.~\eqref{EQ_PT_PP_AV}, finding perfect agreement. We find that 
the $\la P_T^2\ra$ of the produced kaons is significantly larger than that of the pions, 
which is not apparent in the current experimental measurements.  
 
An interesting extension of our model would be to include the vector meson and 
nucleon antinucleon emission channels. 
This extension  has already been completed in our previous work on the integrated fragmentation functions. 
It would also be intriguing to consider the spin-dependent effects in the 
hadronization process, in particular, to calculate the Collins fragmentation function. 
Further, using the NJL description of nucleon structure we will be able 
to develop a self-consistent description of the spin-dependent effects in SIDIS reactions.

 \section*{Acknowledgements}
 
This work was supported by the Australian Research Council through Grants No. FL0992247 
(AWT), No. CE110001004 (CoEPP), and by the University of Adelaide.
\vspace{2em}

\section*{APPENDIX: Nucleon TMD PDF expressions}
\label{sect:nucltmd}
The $u$ and $d$ valence TMD quark distribution functions in the proton 
are given by
\begin{align}
u_v(x,k_T^2) &= f^s_{q/N}(x,k_T^2)  + \tfrac{1}{3}\,f^a_{q/N}(x,k_T^2)  \no \\
&\hs{0mm} + \tfrac{1}{2}\,f^s_{q(D)/N}(x,k_T^2) + \tfrac{5}{6}\,f^a_{q(D)/N}(x,k_T^2),\\[1.0ex]
d_v(x,k_T^2) &= \tfrac{2}{3}\,f^a_{q/N}(x,k_T^2)  \no \\
&\hs{0mm} + \tfrac{1}{2}\,f^s_{q(D)/N}(x,k_T^2) + \tfrac{1}{6}\,f^a_{q(D)/N}(x,k_T^2).
\end{align}
The individual quark diagrams terms have the form
\begin{widetext}
\begin{align}
&f_{q/N}^s(x,k_T^2) =  \frac{\a_1^2\,Z_N\,Z_s}{16\pi^3} \lf(1-x\rg) 
\int d \tau\, 
\lf[ 1 + \tau\,x\lf[\lf(M_N + M\rg)^2 - m_s^2\rg]\rg] e^{-\tau\lf[k_T^2 + x\lf(x-1\rg)M_N^2 + x\,m_s^2 + \lf(1-x\rg)M^2\rg]}, \\
&f_{q/N}^a(x,k_T^2) = -\frac{Z_a\,Z_N}{16\pi^3}\,(1-x)\no \\
&
\int d\tau\,
\lf[\lf(\a_2^2 - 2\a_2\a_3 - 2\a_3^2 \rg) \lf(1 + \tau\,x\,\lf[(M_N - M)^2 - m_a^2\rg]\rg)
- 12 \a_3^2\,\tau\,x\,M\,M_N \rg]
e^{-\tau\lf[k_T^2 + x\lf(x-1\rg)M_N^2 + x\,m_a^2 + \lf(1-x\rg)M^2\rg]},
\end{align}
and the diquark diagrams terms are given by
\begin{align}
f^s_{q(D)/N}(x,k_T^2) &= \iint_0^1 dy \,dz \ \delta\!\lf(x-yz\rg) f^s_{q/D}(z,k_T^2)\ f^s_{q/N}(1-y), \\
f^a_{q(D)/N}(x,k_T^2) &= \iint_0^1 dy\, dz \ \delta\!\lf(x-yz\rg) f^a_{q/D}(z,k_T^2)\ f^a_{q/N}(1-y).
\end{align}
The diquark TMD quark distributions are
\begin{align}
f^s_{q/D}(x,k_T^2) &= \frac{3\,Z_s}{4\pi^3}\, \int d\tau\,\lf[1 + \tau\,x\,(1-x)\,m_s^2\rg]
e^{-\tau\lf[k_T^2 + x(x - 1)\,m_s^2 + M^2\rg]}, \\[1.0ex]
f^a_{q/D}(x,k_T^2) &= \frac{3\,Z_a}{\pi^3}\, x(1-x) 
\int\, d\tau\,
\lf[1 + \tau\,x(1-x)m_a^2\rg] e^{-\tau\lf[k_T^2 + x(x-1)m_a^2 + M^2 \rg]}.
\end{align}
\end{widetext}
The constituent quark, scalar diquark, axial--vector diquark, and nucleon masses in these
expressions have the values $M = 0.4\,$GeV, $m_s = 0.687\,$GeV, $m_a = 1.03\,$GeV, and 
$M_N = 0.94\,$GeV. The weight factors in the nucleon Faddeev vertex~\cite{Mineo:1999eq,Mineo:2002bg,Cloet:2005pp} 
and its normalization are given by
$\lf(\a_1,\,\a_2,\,\a_3\rg) = \lf(0.429,\,0.0244,\,-0.445\rg)$ and $Z_N = 29.9$, respectively. 
Finally the pole residues of the scalar and axial--vector diquark $t$-matrices are 
$Z_s = 14.5$ and $Z_a = 3.82$, respectively. Integrating over $\vect{k_T}$ in these
expressions gives the familiar spin-independent quark distribution functions, the moments
of which satisfy the baryon number and momentum sum rules.

These expressions have been derived using the proper-time regularization scheme, 
which in practice means to make the substitution
\begin{multline}
\frac{1}{X^n} = \frac{1}{(n-1)!}\int_0^{\infty} d\tau\ \tau^{n-1} e^{-\tau X} \\
\longrightarrow \frac{1}{(n-1)!}\int_{1/\Lambda_{UV}^2}^{1/\Lambda_{IR}^2} d\tau\ \tau^{n-1} e^{-\tau X},
\label{eq:pt}
\end{multline}
where $X$ denotes the denominator function in a loop integral after Feynman parameterization and Wick rotation. The
infrared and ultraviolet cutoffs have the values $\Lambda_{IR} = 0.240\,$GeV and $\Lambda_{UV} = 0.645\,$GeV,
respectively.

\bibliographystyle{apsrev}
\bibliography{}

\end{document}